\documentclass[twoside,twocolumn,english]{elsarticle}
\usepackage[T1]{fontenc}
\usepackage[latin9]{inputenc}
\usepackage{amsbsy}
\usepackage{graphicx}
\usepackage{makecell}
\makeatletter
\journal{Example: Nuclear Physics B}
\usepackage[english]{babel}
\usepackage{lineno}

\usepackage{amsmath}
\usepackage{lineno}
\usepackage{hyperref}
\makeatother

\usepackage{babel}
\begin{document}
\twocolumn[   \begin{@twocolumnfalse}

\begin{frontmatter}{}

\title{\textbf{Optimal control of eye-movements during visual search} }

\author{A.Y.Vasilyev}

\address{Queen Mary University of London, Mile End Road, E1 4NS}
\begin{abstract}
We study the problem of an optimal oculomotor control during the execution
of visual search tasks. We introduce a computational model of human
eye movements, which takes into account various constraints of the
human visual and oculomotor systems. In the model, the choice of the
subsequent fixation location is posed as a problem of a stochastic optimal
control, which relies on reinforcement learning methods. We show that
if biological constraints are taken into account, the trajectories
simulated under a learned policy share both basic statistical properties
and a scaling behaviour with human eye movements. We validated our model
simulations with human psychophysical eye-tracking experiments.\end{abstract}

\end{frontmatter}{}

\end{@twocolumnfalse} ]

\section{Introduction}
The human oculomotor system performs hundreds of thousands of eye-movements per
day during the execution of different behavioral tasks. In order to
find the details of a visual scene related to the tasks, humans direct
foveal vision to the most informative locations via saccades - high-velocity conjugate gaze shifts. Saccades are followed by a visual
fixation, during which the human oculomotor system generates fixational
eye movements involuntarily. Despite the remarkable achievements in
the modelling of fixational eye movements and the interpretation of their
fundamental properties \cite{engbert2011integrated,engbert2006microsaccades,engbert2004microsaccades},
there is no comprehensive generic model of fixation selection \cite{tatler2011eye,borji2013state,ballard2009modelling},
which takes into account the underlying mechanisms of visual attention
\cite{borji2013state,ko2010microsaccades,sinn2015small} and qualitatively
describes the statistical properties of saccadic eye-movements during the
execution of visual tasks \cite{najemnik2009simple,amor2016persistence,najemnik2005optimal,stephen2011fractal}. 

Previously, the problem of fixation selection was studied in the framework
of control models of eye movements \cite{butko2010infomax,najemnik2005optimal,najemnik2009simple}.
In control models the observer gathers information about the world
during each fixation, integrates information over all fixations into a belief state and
makes a choice of the next location on which to fixate. This choice
is governed by the policy of gaze allocation - a function that specifies the  action of decision-maker in a certain belief state. It was
shown that the policy based on information maximization criteria \cite{najemnik2009simple}
generates trajectories that share basic statistical properties with
human eye movements. In this research, we set the goal of developing a
control model of fixation selection that is capable of interpreting
the scaling behaviour of human eye-movements \cite{amor2016persistence,wallot2015cue,stephen2011fractal,grigolini2009theory}
and provides a human level of performance to a computational agent. 

In contrast to the previous research on control models, we take into account
the inherent uncertainty of human oculomotor system and the duration of saccadic
eye movements. It's well known that any motor action of humans is executed
with random error, which increases with movement magnitude \cite{van2007sources,engbert2005swift}.
Despite the oculomotor system having developed a correction mechanism
for saccade errors \cite{tian2013revisiting}, these result in inevitable
temporal costs. Furthermore, the duration of saccades is empirically
correlated with their magnitude as well \cite{lee2002eyes}. These
factors result in situations where the observer has to choose between
more informative remote (and riskier) locations and those nearby (but
less informative ones). We show that if these constraints are taken
into consideration, the trajectories simulated under a learned policy
share both basic statistical properties and scaling behavior with
human eye movements, which is not achievable with the conventional infomax
model \cite{najemnik2009simple}. 
\par On the basis of our results, we argue that we have made the following
contribution:
\begin{itemize}
\item The formulation of the biologically plausible model of gaze allocation in
the human observer from the point of view of stochastic optimal control.
The representation of the model in the form of partially observable Markov
decision process (PO-MDP) and the proposal of a heuristic policy. 
\item The development of robust and high performance algorithms of simulation
of PO-MDP. The implementation of reinforcement learning algorithms of
policy optimization and numerical estimation of the optimal policy of
gaze allocation. 
\item The comprehensive statistical analysis of simulated trajectories and data
from our psycho-physical experiments. The policy, which is learned
with the policy gradient REINFORCE algorithm, shows the highest level
of statistical similarity with human eye-movements. In our experiments
we discovered the dependency of the mean saccade length and q-order Hurst
exponent on visibility of the target, which was explained by our model.
\end{itemize}

\section{Model of the ideal observer}

\begin{figure}
\centering
\includegraphics[scale=0.75]{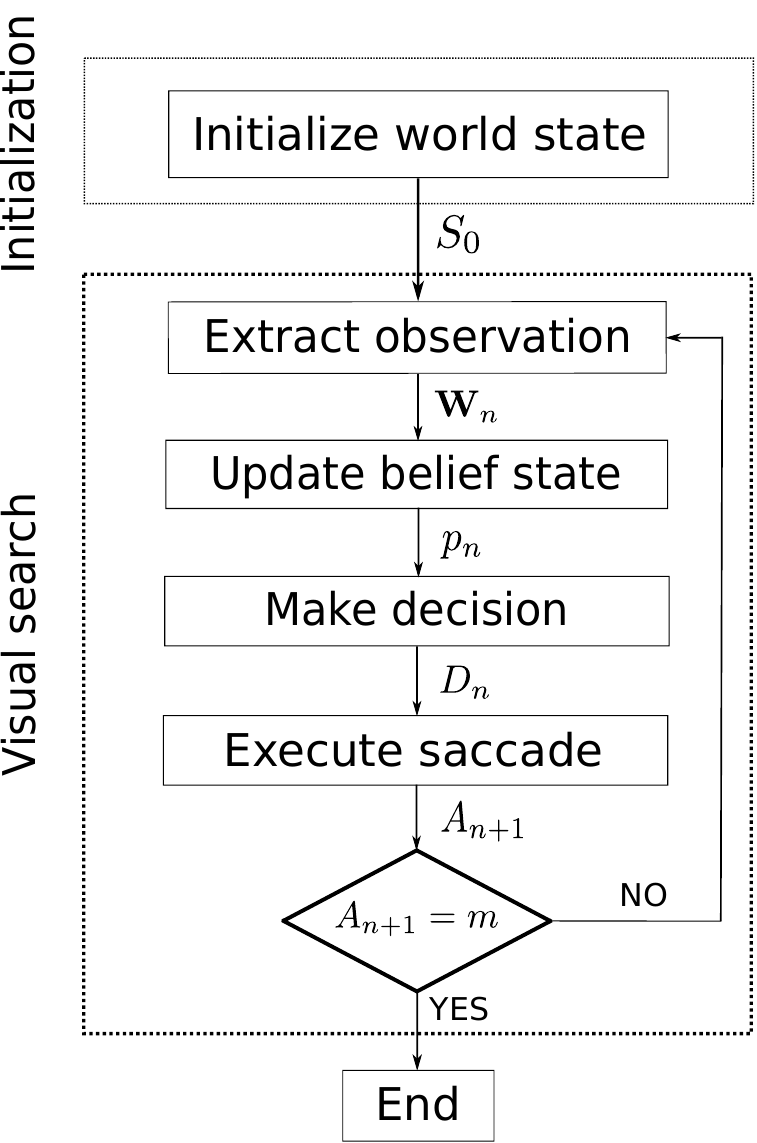}\caption{\label{fig:flowchart} Flow chart of the model of the ideal observer. The visual search is a recurrent process that starts after initialization of the world state. On each consequent step the observer receives the observation vector $\mathbf{W}_{n}$, which is then used for estimation of belief state $p_{n}$ using Bayesian inference \cite{butko2008pomdp}. After the update of the belief state the observer makes the decision $D_{n}$ where to fixate next according to the policy of gaze allocation: $D_{n}=\mu(p_{n})$. The next fixation location is defined by execution function: $A_{n+1}=\alpha(D_{n}) $. If the observer fixates on the location of target $A_{n+1}=m$ the process of visual search is terminated, otherwise the next step starts with updated values of variables. }
\end{figure}

In this section we formulate the model of the ideal observer, which aims
to localize the single target object on the stationary 2D image. We represent the model in the form of partially observable Markov
decision process (PO-MDP), which is summarized by flow chart on Figure  \ref{fig:flowchart}. 

\subsection{World state}

At the beginning of each episode the target object appears randomly
at one of $L$ possible locations. We assume that the target is
placed on background noise or surrounded by distractors, which are
placed on vacant locations. The world state $S_{n}$ is represented
as a tuple:

\begin{equation}
S_{n}=\left(m,\ A_{n},\ t_{n}\right)\label{eq:worldstate}
\end{equation}
where $m$ is a location of the target on the image and $A_{n}$ is
gaze fixation location that changes with the number of step $n$,
and $t_{n}$ is time passed from the start of a trial and the step
$n$. 

If the observer fixates the gaze on the location of target:
\begin{equation}
A_{n}=m\label{eq:terminal}
\end{equation}
the visual task is considered to be accomplished. This formulation of the terminal state reflects the necessity to foveate the target in
order to extract as much information about its identity and details
as possible.  The location of target $m$ doesn't change during a trial.

\subsection{Update of belief state }
The decision-making of the observer is modeled as PO-MDP with a belief
state $p_{n}$ - a discreet probability distribution function of target location
given all observations received up to the step $n$.  Because the observer is instructed that the  target appears randomly, the initial belief state $p_0$ is a discrete uniform distribution.
\par
\begin{figure}
\centering
\includegraphics[scale=0.55]{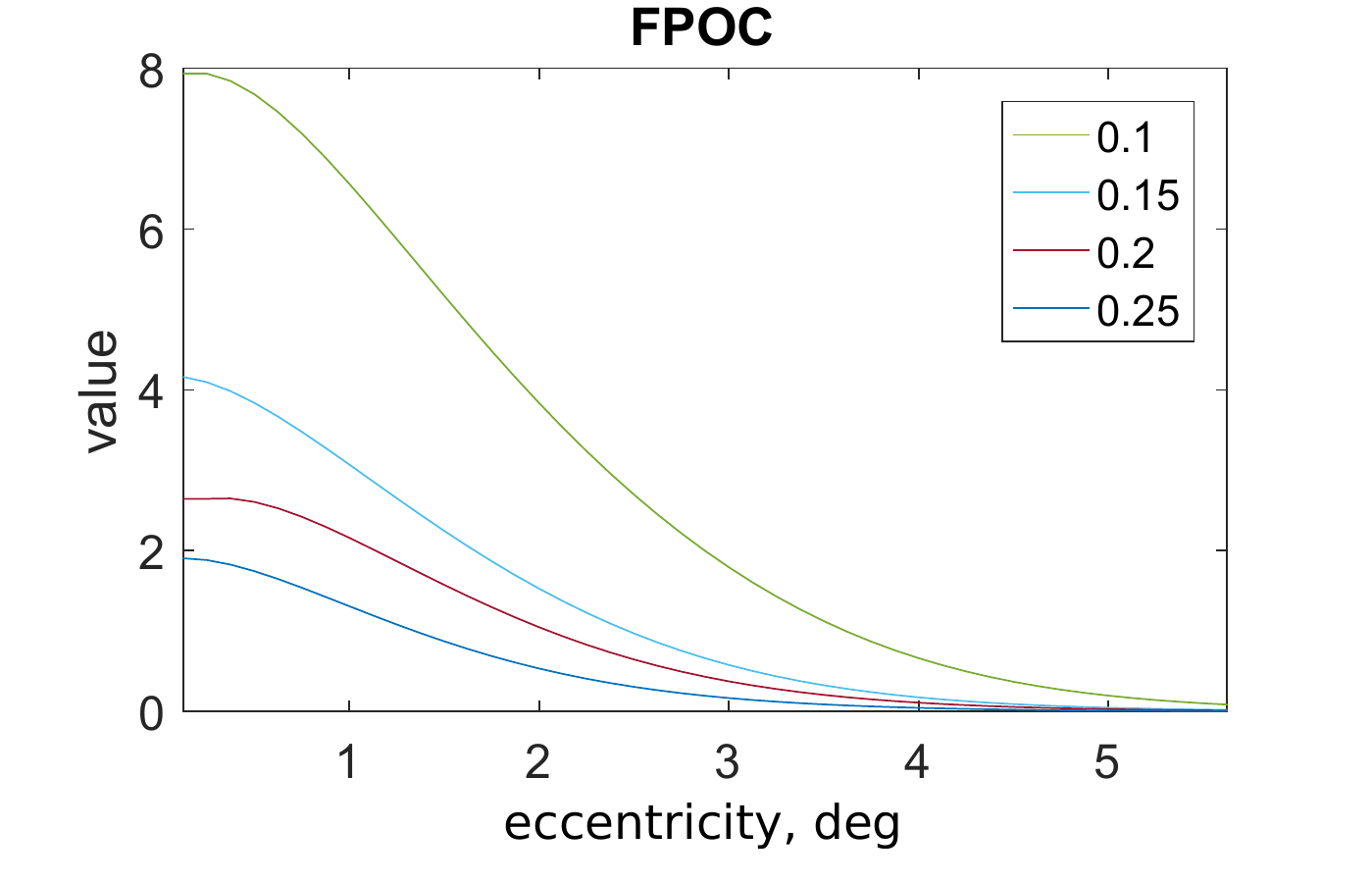}\caption{\label{fig:Fovea-Peripheral-Operating-Characteristics}Fovea-Peripheral Operating
Characteristic was calculated for several values of the RMS contrast of
background noise: $e_{n}\in\left(0.1,0.15,0.2,0.25\right)$ and the single
value of RMS contrast of target $e_{t}=0.2$. The signal-to-noise ratio
has a peak at fovea and decreases rapidly with eccentricity. }
\end{figure}

 On each step $n $ observer receives the observation vector $\mathbf{W}_{n}=\left(W_{1,n},\ldots,W_{L,n}\right)$,
whose elements represent the perceptual evidence that the target is
at corresponding locations. The probability distribution function
 is updated using Bayesian inference \cite{butko2008pomdp}:

\begin{equation}
\mathrm{\mathit{p_{n}(l)=\frac{p_{n-1}(l)p(\mathbf{W}_{n}|l,A_{n})}{\underset{k}{\sum}p_{n-1}(k)p(\mathbf{W}_{n}|k,A_{n})}}}\label{eq:inference}
\end{equation}
where $l$ is the index of the location and $p(\mathbf{W}|l,A)$ is
an observation model. In order to take into account the uncertainty
of the processing of perceptual information within the neural circuits
of the observer, we follow the ``noisy observation'' paradigm \cite{najemnik2009simple}.
In this paradigm the observation model $p(\mathbf{W}|l,A)$
reflects the presence of the observer's internal sources of inefficiency,
such as physical neural noise on all stages of information processing.
According to the perceptual model \cite{najemnik2005optimal} the observation
$\mathbf{W}$ may be represented as a random variable with Gaussian
distribution with mean depending  on the location $m$ of the center of target
on the lattice:

\begin{align*}
p(\mathbf{W}|l,A)=\underset{l}{\prod}p(W_{l}|A) & =
\end{align*}

\begin{equation}
\underset{l}{\prod}N\Bigl(W_{l};\delta_{l,m}^{\mathbf{}},\frac{1}{F\left(\left\Vert l-A\right\Vert \right)}\Bigr)\label{eq:observation}
\end{equation}
where $\delta_{i,j}^{\mathbf{}}$ is Kronecker delta, $N(x,\mu,\nu)$
is a value of Gaussian function with mean $\mu$ and variance $\nu$
for argument $x$; $\left\Vert l-A\right\Vert $ is Euclidean
distance between the locations $l$ and the current fixation $A$, and
$F$ is a Fovea-Peripheral Operating Characteristic (FPOC) \cite{butko2010infomax}.
FPOC is a function that represents the dependence of a signal-to-noise ratio
on the eccentricity. Figure  \ref{fig:Fovea-Peripheral-Operating-Characteristics}
demonstrates FPOC calculated for several values of RMS contrast of
the background 1/f noise: $e_{n}\in\left(0.1,0.15,0.2,0.25\right)$ and a
single value of RMS contrast of target $e_{t}=0.2$. The calculation
are based on the analytical expressions from \cite{najemnik2005optimal}.
The signal-to-noise ratio has a peak at fovea and decreases rapidly with
eccentricity.
\par In our simulations we consider only the case of the rotationally
symmetric FPOC. This assumption is not correct for human observers, and better generic model of FPOC can be found in \cite{bradley2014retina}. The broken circular symmetry of FPOC inevitably results in the asymmetry of the visual search process \cite{najemnik2008eye}. We simplify the model of FPOC, because in this research we focus our attention more on the temporal structure of eye-movements rather than on their spatial distribution. 

\subsection{Execution of saccades}

The decision of which
location to fixate next, $D_{n}$, is made on each step of PO-MDP
according to the policy of gaze allocation $\mu:$
\begin{equation} D_{n}=\mu(p_{n}) \label{eq:decisionmake}
\end{equation}
After making the decision, the coordinates of the next fixation location $A_{n+1}$
are defined by the execution function: 
\begin{equation}
A_{n+1}=\alpha(D_{n})=D_{n}+J_{n}\label{eq:saccade-1}
\end{equation}
where $J_{n}$ is a Gaussian-distributed random error with zero mean
and standard deviation $\nu$ defined in \cite{engbert2005swift}:

\begin{equation}
\nu=\zeta_{0}+\zeta_{1}\left\Vert D_{n}-A_{n}\right\Vert \label{eq:saccadevar-1}
\end{equation}

The error of the saccade execution is proportional to intended saccade
amplitude $\left\Vert D_{n}-A_{n}\right\Vert $ given in degrees,
the value of parameters: $\zeta_{0}=0.87\deg$, $\zeta_{1}=0.084$  (from\cite{engbert2005swift}). 

The next step of PO-MDP starts after the transition to the location $A_{n+1}$.  This decision-making model may be easily extended in order to take into account the extraction of visual information between the moment of making  the decision where to fixate next and completion of the saccade.  

\subsection{Duration of the steps}

After each consequent step the time variable $t$ of world state (\ref{eq:worldstate})
is updated in a deterministic way:

\begin{equation}
t_{n}= t_{n-1}+\Theta(n)\label{eq:time_update}
\end{equation}
where $\Theta(n)$ is a duration of step $n$. The duration of time
step $\varTheta(n)$ is considered as a total time, which is required
for the relocation of the gaze from a previous location $A(n-1)$ to the current one $A(n)$ and the extraction of visual information from
the location $A(n)$. Therefore, we consider $\varTheta(n)$ as a sum of durations of the fixation
$\varTheta_{fix}(n)$ and the saccade $\varTheta_{sac}(n)$.
According to the literature, both of these time intervals are empirically
correlated with a magnitude of the saccade preceding the fixation \cite{lee2002eyes,bartz1962eyemovement,salthouse1980determinants}.
The duration of saccadic eye-movements $\varTheta_{sac}(n)$ in range
of magnitudes from $1.5\textdegree$ to $30\textdegree$ is possible
to approximate as \cite{lebedev1996square}:

\begin{equation}
\varTheta_{sac}(n)=\tau_{sac}\left\Vert A_{n}-A_{n-1}\right\Vert ^{0.4}\label{eq:duration_sac}
\end{equation}
where $\tau_{sac}=21 ms\cdot \deg^{-0.4}$. Besides the magnitude
of saccade, the fixation duration $\varTheta_{fix}(n)$ is influenced
by various factors as a discriminability of the target \cite{hooge1996control},
its complexity and the visual task of the observer \cite{salthouse1980determinants,kliegl2006tracking}.
However, if the observer is correctly informed about the targets'
properties before the task execution and performs the visual task
without any interruptions, the contribution of these factors to the fixation
duration (with exception of magnitude) is constant during each trial.
The eye-tracking experiments with the fixations tasks \cite{bartz1962eyemovement,salthouse1980determinants,rayner1978eye}
found that the dependence of fixation duration on saccade amplitude is
linear:

\begin{equation}
\varTheta_{fix}(n)=\left\Vert A_{n}-A_{n-1}\right\Vert \tau_{fix}+\varTheta_{0,fix}\label{eq:duration_fix}
\end{equation}
with a slope $\tau_{fix}=6 ms/\deg$. The constant $\varTheta_{0,fix}=250ms$ is an intercept,
averaged from values from eye-tracking data \cite{greene2006control,unema2005time}.
Finally, the duration of step $n$ is:

\begin{equation}
\varTheta(n)=\varTheta_{sac}(n)+\varTheta_{fix}(n)\label{eq:duration}
\end{equation}
The values of parameters used in simulations are consistent with our estimates from the eye-tracking experiments: $\tau^{*}_{sac}=20 \pm 3 ms \cdot  \deg^{-0.4}$,  $\tau^{*}_{fix}=5.8 \pm 1.8  ms/\deg$,  $\varTheta_{0,fix}=241 \pm 42 ms$.  Within this range of the parameters' values we didn't find a substantial difference in the estimates of the learned policy of gaze allocation.

\subsection{Value function}

Given the initial world state $S_{0}$, we define
the cost function for policy $\mu$ as an expectation of a random variable
$V$: 

\begin{equation}
V_{\mu}(S_{0})=E\left[V|\mu,S_{0}\right]\label{eq:value_function}
\end{equation}
The random variable $V$ denotes the cost and is defined by:

\begin{equation}
V\equiv c\sum_{n=0}^{N}\varTheta(n)=ct_{N}\label{eq:reward}
\end{equation}
where $N$ is a total number of steps in the episode, and $c$ is a time
cost constant. 

Formulation of the cost function in a real time sets this study separately
from the previous works \cite{navalpakkam2010optimal,najemnik2005optimal,butko2010infomax}.
We show below, that the policy $\mu$ optimized for the cost function
with the reward defined in (\ref{eq:reward}) generates the sequences of actions with statistical characteristics close to the human saccadic eye-movements. 

\section{Policy of gaze allocation}

\subsection{Infomax approach\label{sub:heuristics}}

In this section we describe two heuristic policies related to the
model of Entropy Limit Minimization searcher \cite{najemnik2009simple}.
We define the information gain on the step $n+1$ as: $\triangle I(n+1)=H(p_{n})-H(p_{n+1})$,
where $H\left(\cdot\right)$ is Shannon entropy. The heuristic policy
$\pi_{0}$ is defined as a policy which chooses such decision $D_{n}$
that maximizes the expected information gain $\triangle I(n+1)$:

\begin{equation}
\pi_{0}(p_{n})=D_{n}=\underset{D}{\arg\max}\left[E\left[\triangle I(n+1)\right]\right]
\end{equation}
The term $E\left[\triangle I(n+1)\right]$ is calculated analytically
in \cite{najemnik2009simple} for the case of the saccadic eye-movements
without uncertainty ($A_{n+1}\equiv D_{n})$:

\begin{equation}
E\left[\triangle I(n+1)\right]=\frac{1}{2}\left(p_{n}*F^{2}\right)\left(D_{n}\right)\label{eq:pi0}
\end{equation}
where sign $*$ denotes a convolution operator, and $F$ is FPOC represented
as a radially symmetric 2D function: $F(A)\equiv F\left(\left\Vert A\right\Vert \right)$.
The expression (\ref{eq:pi0}) gives an approximate value of the expected
information gain in the case of the stochastic saccadic placement (\ref{eq:saccade-1}). 

The figure \ref{fig:Where-to-fixate} illustrates the decision-making
process, which corresponds to the policy $\pi_{0}$. The colour map (left)
represents the function of the expected information gain (equation (\ref{eq:pi0})).
The blue cross corresponds to the  location of the current fixation on
the step $n$. The observer makes a decision to fixate at the location defined
by the policy: $D_{n}=\pi_{0}(p_{n})$. This decision results in a saccadic
eye-movements to location $A_{n+1}=\alpha(D_{n})$ marked by the green
cross. After receiving the observation at the step $n+1$, the observer updates the
belief state and evaluate the information gain for the next decision.
In this particular situation, the target was absent at the vicinity
of $A_{n+1}$, and the observation resulted in the decline of probability $p_{n+1}$
in the area around the green cross (figure \ref{fig:Where-to-fixate}
right). This area is effectively inhibited from subsequent fixations
due to low probability. The size of this area is defined by values
of FPOC ($e_{t}=0.2$, $e_{n}=0.1$ in this case). We call the policy $\pi_{0}$ ``infomax greedy'' in the text below.

The trajectories generated with infomax greedy policy match the basic properties
of human eye movements \cite{najemnik2009simple}. However, the policy
(\ref{eq:pi0}) doesn't consider the correlation between the magnitude
of saccades and the durations of steps of MDP. We show later that
the policy $\pi_{0}$ is inferior to the policy that optimizes the
expected rate of information gain $E\left[\triangle I(n+1)/\varTheta(n+1)\right]$:

\begin{equation}
\pi_{1}(p_{n})=\underset{D}{\arg\max}\left[E\left[\triangle I(n+1)/\varTheta(n+1)\right]\right]
\end{equation}

Using the expression for $E[\triangle I(n+1)]$ (\ref{eq:pi0}), for the
deterministic saccadic placement $(A_{n+1}\equiv D_{n})$:

\begin{figure*}

\hspace*{\fill}\includegraphics[scale=0.47]{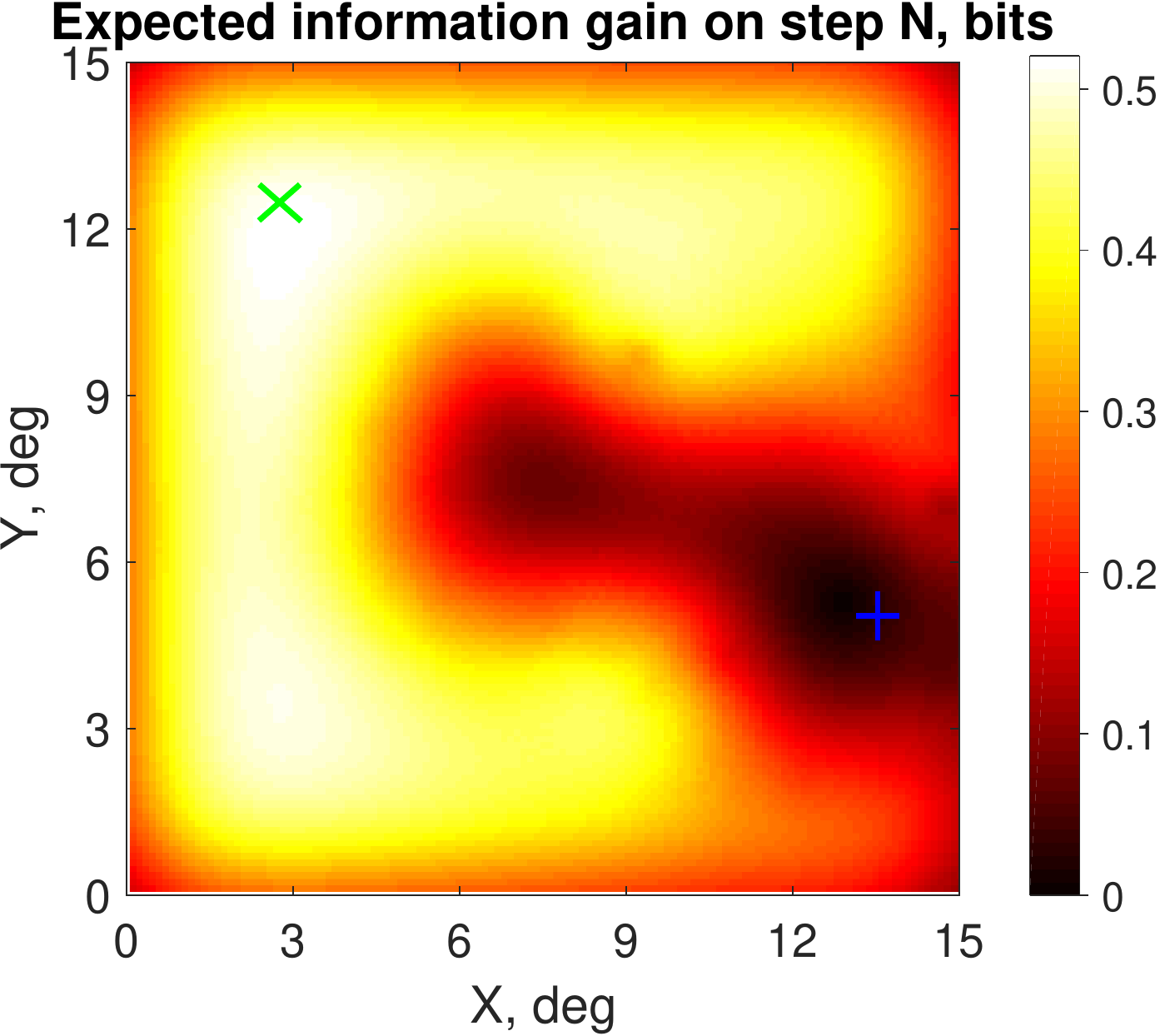}\hfill\includegraphics[scale=0.47]{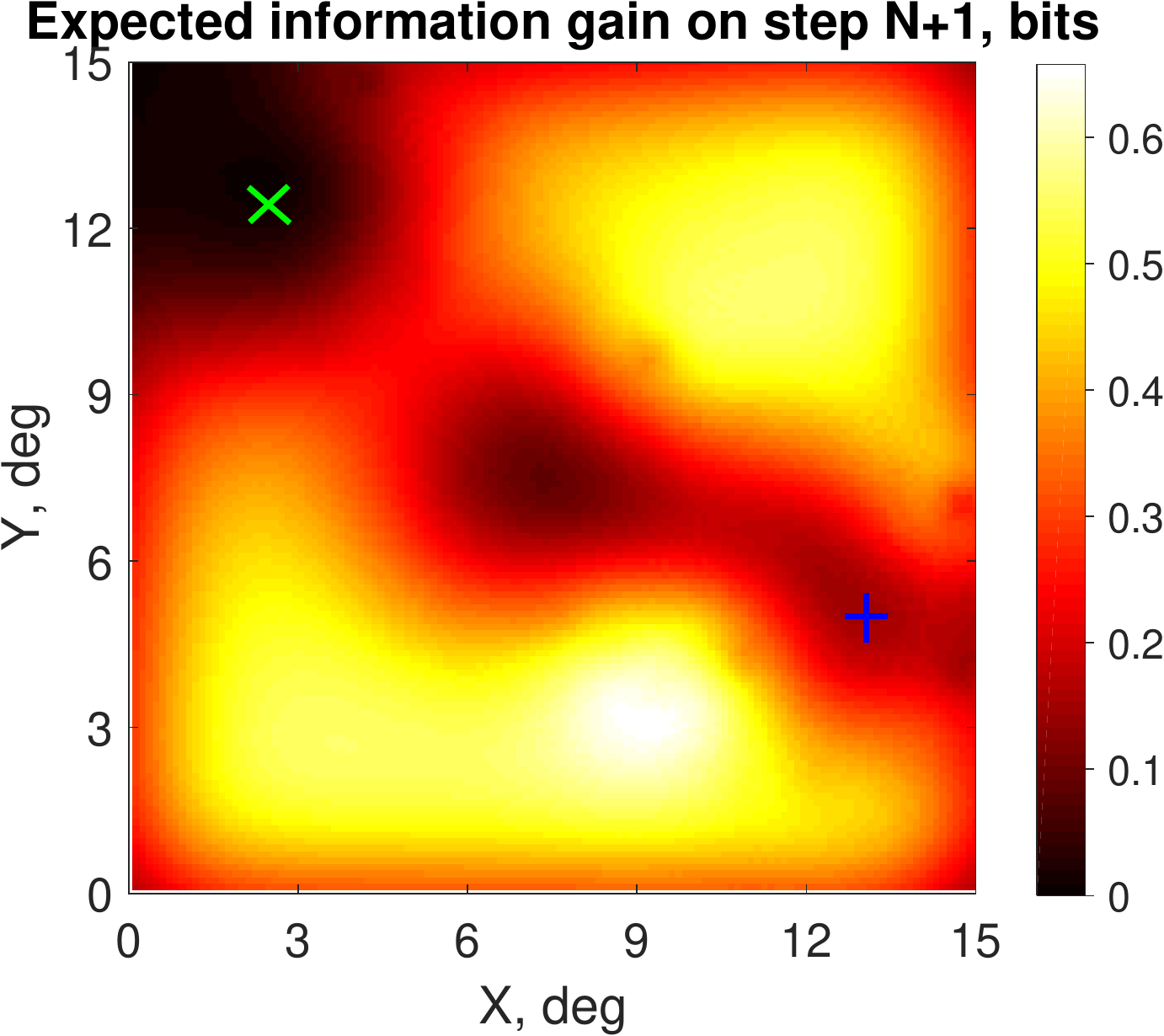}\hspace*{\fill}\caption{\label{fig:Where-to-fixate}The decision-making process under the infomax
greedy policy $\pi_{0}$ \cite{najemnik2009simple}. The colour map
(left) represents the function of the expected information gain (equation
(\ref{eq:pi0})). The blue cross corresponds to location of the current
fixation on the step $n$. The observer makes a decision to fixate at
the location defined by the policy: $D_{n}=\pi_{0}(p_{n})$. This decision
results in the saccadic eye-movement to location $A_{n+1}=\alpha(D_{n})$
marked by the green cross. After receiving the observation at the step $n+1$,
observer updates the belief state and evaluates the information gain for the
next decision. In this particular situation, the target is absent
in the vicinity of $A_{n+1}$, and the observation resulted in the decline
of probability $p_{n+1}$ in the area around the fixation (the green cross).
This area is effectively inhibited from the subsequent fixations due to
low probability $p_{n+1}$. The size of this area is defined by the values
of FPOC (in this case $e_{t}=0.2$, $e_{n}=0.1$).}
\end{figure*}

\begin{equation}
\pi_{1}(p_{n})=D_{n}=\underset{D}{\arg\max}\left[\frac{\left(p_{n}*F^{2}\right)\left(D_{n}\right)}{\varTheta(n+1)}\right]\label{eq:pi_0-1}
\end{equation}

The policy $\pi_{1}$ is called "infomax rate'' in the text below.
The performance of these two heuristic policies will be compared with
a performance of the policy learned with reinforcement learning algorithms
in the section \ref{sub:Convergence-of-policy-1}.

\subsection{Optimal policy estimation}

In this section we describe the evaluation of the policy of gaze allocation that optimizes the cost function (\ref{eq:value_function}) for any starting world state $S_{0}$.
We start with the representation of the stochastic policy $\mu$ in the
on \cite{butko2008pomdp}:

\begin{equation}
\mu(D,p)=\frac{\exp(f(D,p))}{\underset{l}{\sum}\exp(f(l,p))}\label{eq:soft_max}
\end{equation}
where $f(D,p)$ is a function of expected reward gain after making
the decision $D$ with the belief state $p$. In this study we limit
the search of $f(D,p)$ to a convolution \cite{butko2008pomdp}
of belief state $p$:

\begin{equation}
f(D,p)=\underset{l}{\sum}K(D-l)p(l)\label{eq:gainfunction}
\end{equation}

\begin{figure*}

\hspace*{\fill}\includegraphics[scale=0.48]{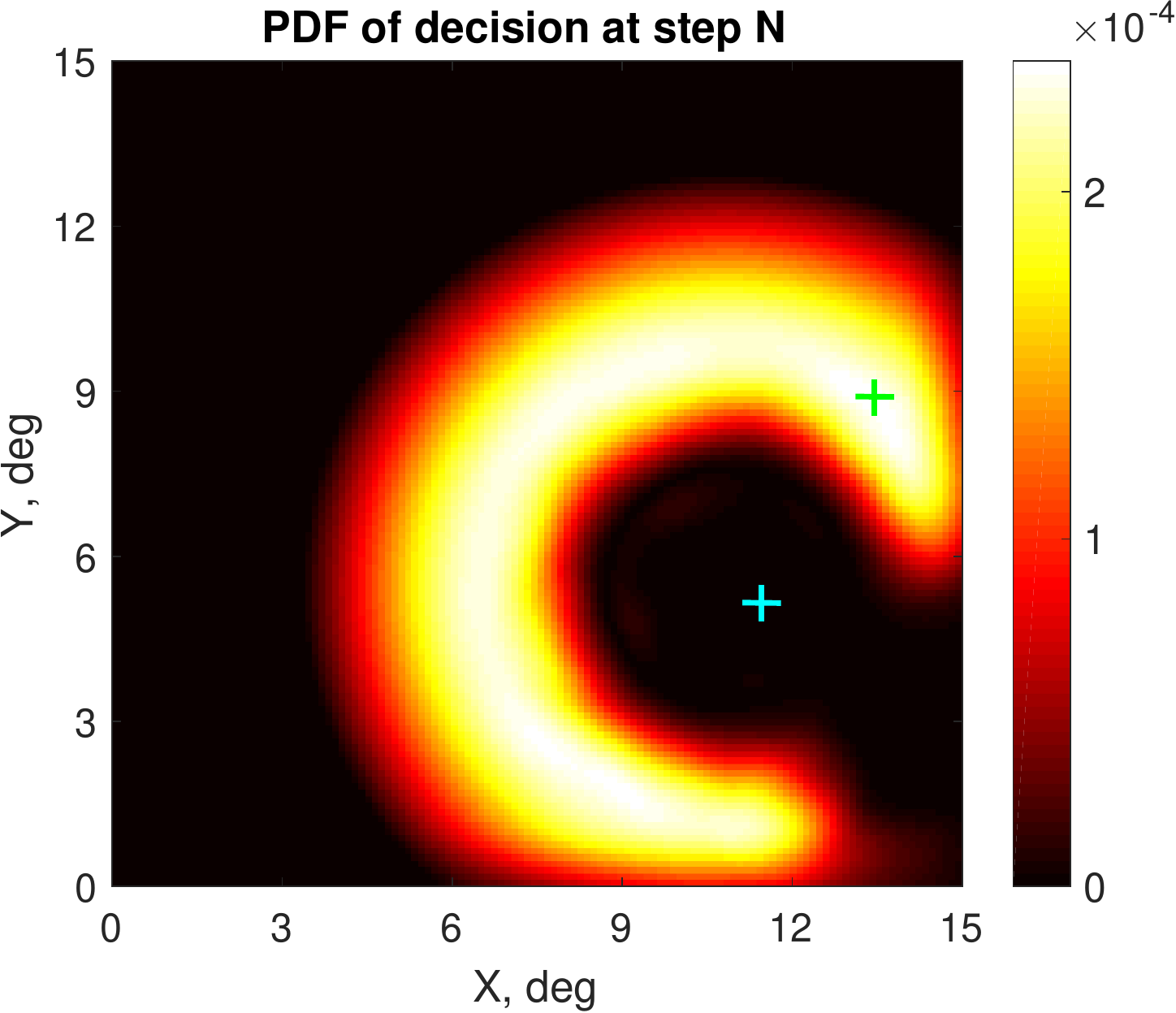}\hfill\includegraphics[scale=0.48]{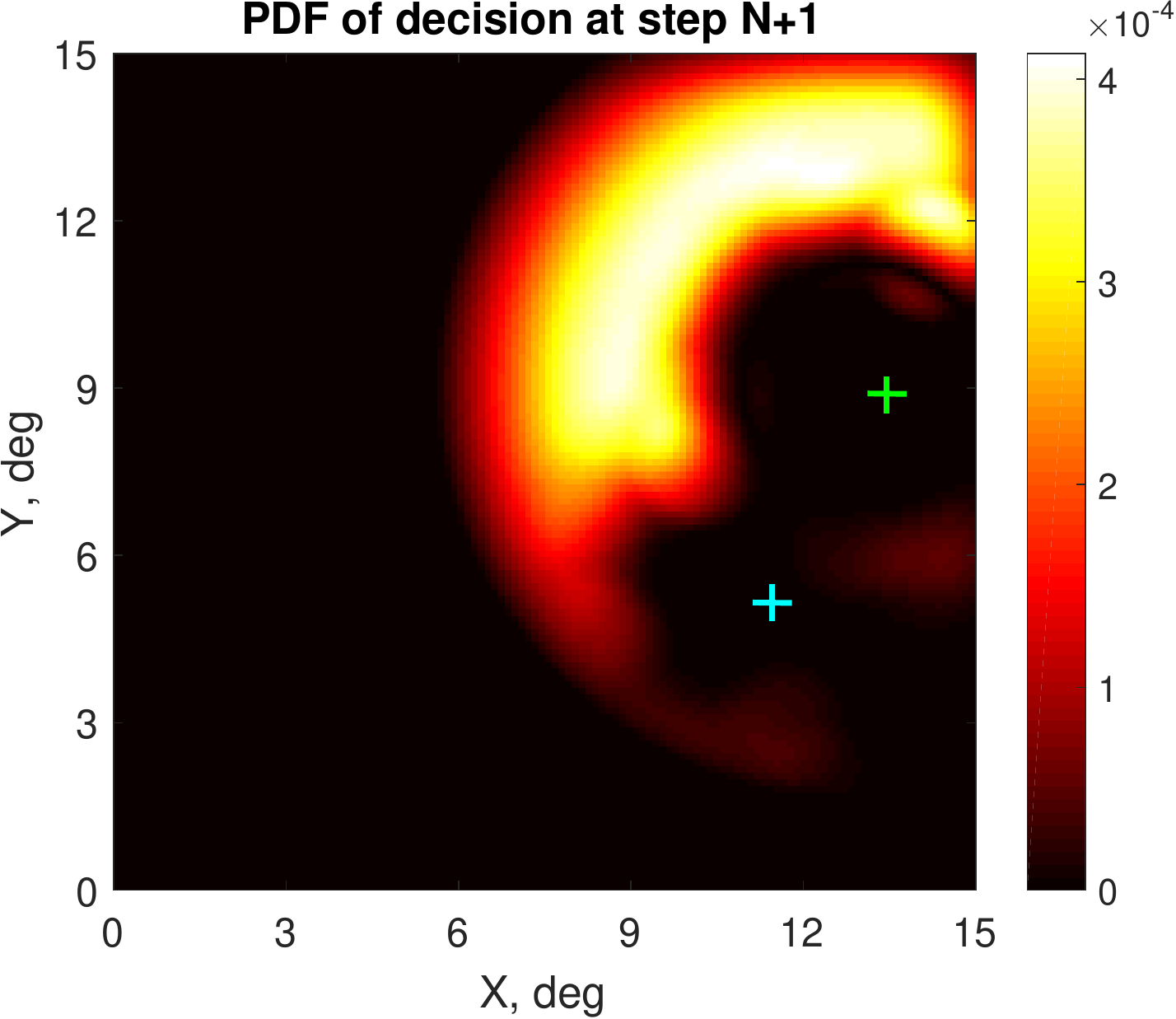}\hspace*{\fill}\caption[The decision-making process under the learned policy.
]{\label{fig:Where-to-fixate-1}The decision-making process under the policy
learned for FPOC corresponding to conditions $e_{t}=0.2$, $e_{n}=0.1$.
At the step $n$ observer fixates the location marked by a blue cross.
The policy $\mu$ defines a probability density function of a decision
$D$ where to fixate next (\ref{eq:policyform}). Observer chooses
the decision $D_{n}$ according to the policy, which results in a saccadic
eye-movement to location $A_{n+1}=\alpha(D_{n})$ (the green cross).
As well as in the case of dynamics under the heuristic policy $\pi_{0}$
previously visited locations are inhibited from the subsequent fixations.
Note that movements to remote locations are inhibited by the radial function.
This results in co-directed short movements, which are also characteristic
of human observer.}

\end{figure*}
\par  In supplementary materials \ref{sub:separability}  we justify this choice of the policy and evaluate the form of a kernel function $K$ that allows us to effectively solve the optimization problem with the policy gradient algorithms. Our task is the search of the kernel function $K$
(\ref{eq:separability}), which
corresponds to the policy that optimizes the cost function $V_{\mu}$:

\begin{equation}
K^{*}=\underset{K}{\arg\min}\:V_{\mu(K)}(S_{0})\label{eq:problem}
\end{equation}
for any starting world state $\forall S_{0}$. The policy $\mu(K^{*})$
is called the optimal policy of gaze allocation. \par  We approach the optimization problem (\ref{eq:problem})  with an algorithm named "REINFORCE with optimal
baseline'' \cite{peters2006policy} according to the procedure described
in Supplementary material\ref{sec:Implementation-of-reinforcement}. The performance of
REINFORCE was compared with one of the optimization algorithms named
"policy gradient parameter exploration'' (PGPE) adopted from \cite{sehnke2010parameter}.
The algorithm of REINFORCE with an optimal baseline belongs to the class
of the likelihood ratio methods, whereas PGPE is related to the finite difference
methods. Despite the distinction between these two approaches, both
algorithms give a close estimation of the optimal policy \ref{sub:Convergence-of-policy-1}.
We simulated trajectories for the data analysis in section \ref{sec:Analysis-of-data}
using the solution provided by REINFORCE due its better performance
comparing to PGPE. 
Figure \ref{fig:Where-to-fixate-1} demonstrates the decision-making
process under the policy $\mu$ learned for FPOC corresponding to the conditions
$e_{t}=0.2$, $e_{n}=0.1$ (see figure \ref{fig:Results-of-optimization:-1}
for its kernel function). At the step $n$ the observer fixates the location
marked by a blue cross. The policy $\mu$ defines a probability density
function of the decision $D$ where to fixate next (\ref{eq:policyform}).
The observer chooses the decision $D_{n}=\mu\left(p_{n}\right)$, which
results in a saccadic eye-movement to the location $A_{n+1}=\alpha(D_{n})$
(green cross). As well as in the case of dynamics under the heuristic
policy $\pi_{0}$, previously visited locations are inhibited from the 
subsequent fixations.  

\section{Basic properties of trajectories\label{sec:Analysis-of-data}}

In this section we discuss the statistical properties of trajectories
generated with the learned policy $\mu$ and the heuristic policies $\pi_{0}$
and $\pi_{1}$. The simulations were performed on the grid with size $128\times128$
that corresponds to the visual field with size of $15\times15\deg$
in the psychophysical experiment. In order to justify our computational model, we reproduced the psychophysical experiments from \cite{najemnik2009simple}. The detailed description of the experiments can be found in Appendix\ref{sec:Qualitative-analysis}.

\subsection{Performance}

Although this computational model was not designed for an exact prediction
of a response time of human observers, it demonstrates a high level of
consistency in a performance of the visual task execution with human observers. The performance
was measured as an average time to reach the target (the mean response time)
and as a percentage of the correct fixations on target's location on an
N-Alternative Forced Choice task (N-AFC). The unsuccessful trials from the
psychophysical experiments were excluded from the consideration.  We found that the number of the unsuccessful trials grows with the contrast of noise: $2.3\%$, $5.7\%$, $9.8\%$, $16.4\%$  for the corresponding numbers of the contrast $\epsilon_{n}=$ $(0.1$, $0.15$, $0.2$, $0.25)$.

\begin{figure*}
\hspace*{\fill}\includegraphics[scale=0.56]{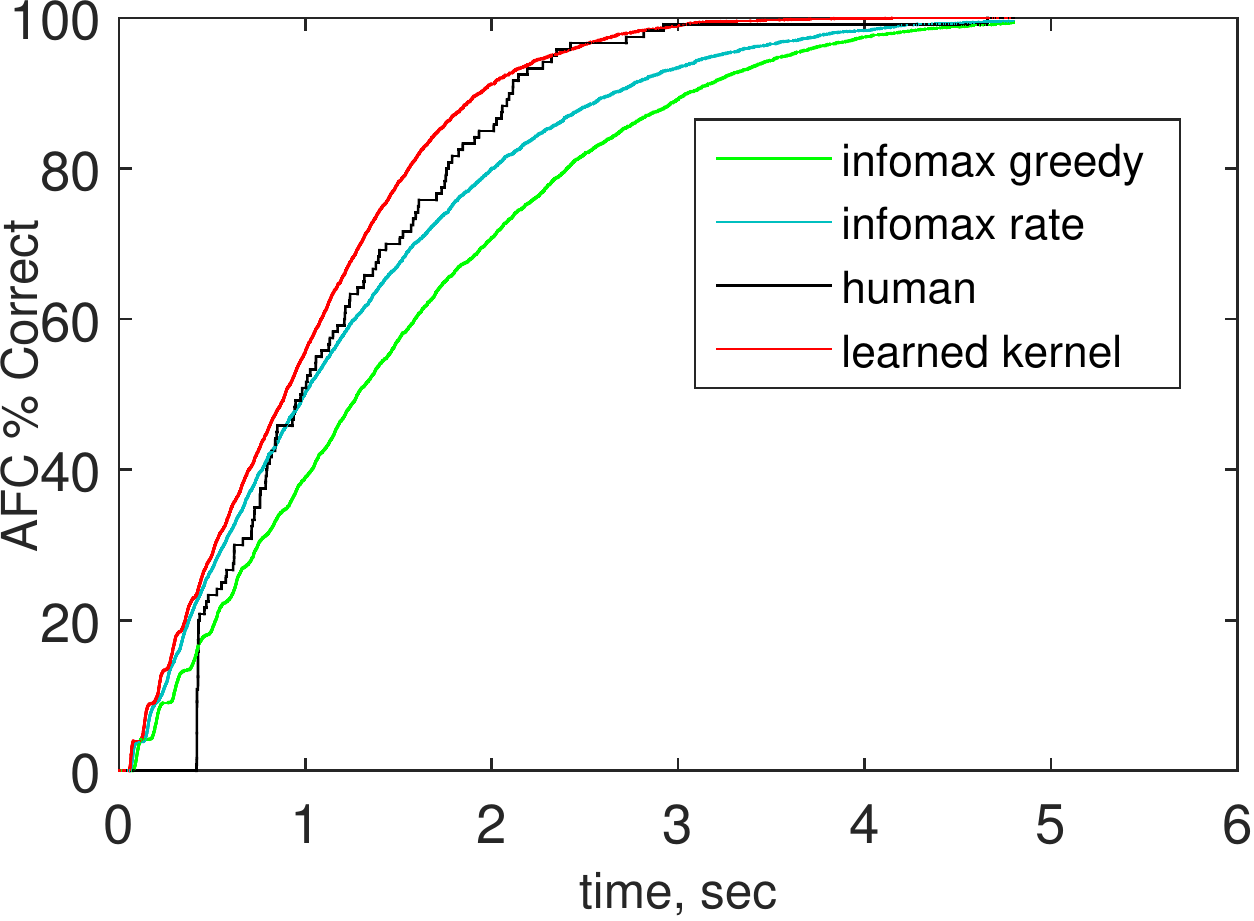}\hfill\includegraphics[scale=0.47]{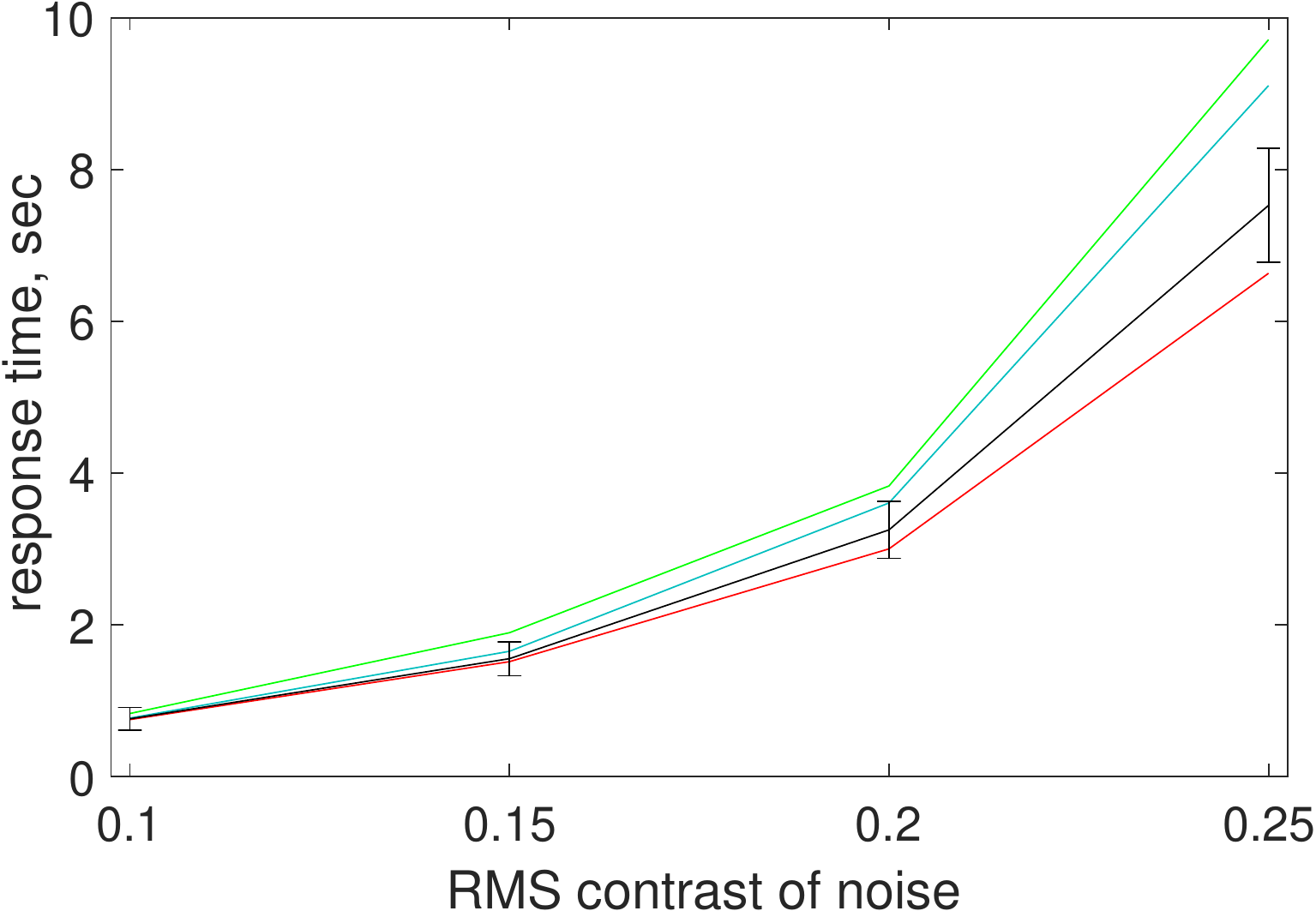}\hspace*{\fill}\caption[The performance of the human observers and
the simulated agents.]{\label{fig:Performance-of-human}The performance of the human observers and
the simulated agents. The learned policy outperforms two heuristics both
in the mean completion time and the percentage of correct responses in N-AFC
task (left) for the experimental conditions: $e_{t}=0.2$, $e_{n}=0.15$. The dependence of mean completion time (right) for the learned
policy resembles one for the human observer.}

\end{figure*}

Figure \ref{fig:Performance-of-human} (left) demonstrates the percentage
of correct fixations on the target location for the experimental conditions:
$e_{t}=0.2$, $e_{n}=0.15$. Means and standard errors of the response time of the human observers is presented on Figure \ref{fig:Performance-of-human} (right) together with means of the response time for three policies estimated from $10^{4}$ episodes of PO-MDP.  The learned policy outperforms two heuristics and the human observers
both in the mean response time and the percentage of the correct fixations for all experimental conditions. Human observers significantly outperformed the infomax rate for the experimental conditions: $e_{n}=(0.2, 0.25)$ (Student's t-test $p<0.05$) and the infomax greedy for the conditions $e_{n}=(0.15, 0.2, 0.25)$ ($p<0.05$) on the mean response time, which was previously found in \cite{najemnik2008eye,najemnik2009simple}. In the same time the learned policy outperformed the human observers significantly for the condition $e_{n}=0.25$, while for other conditions t-test didn't reject hypothesis that distributions have equal means at $5\%$ significance level. 

\subsection{Amplitude distribution\label{sub:Amplitude-distribution}}

\begin{figure*}
\hspace*{\fill}\includegraphics[scale=0.56]{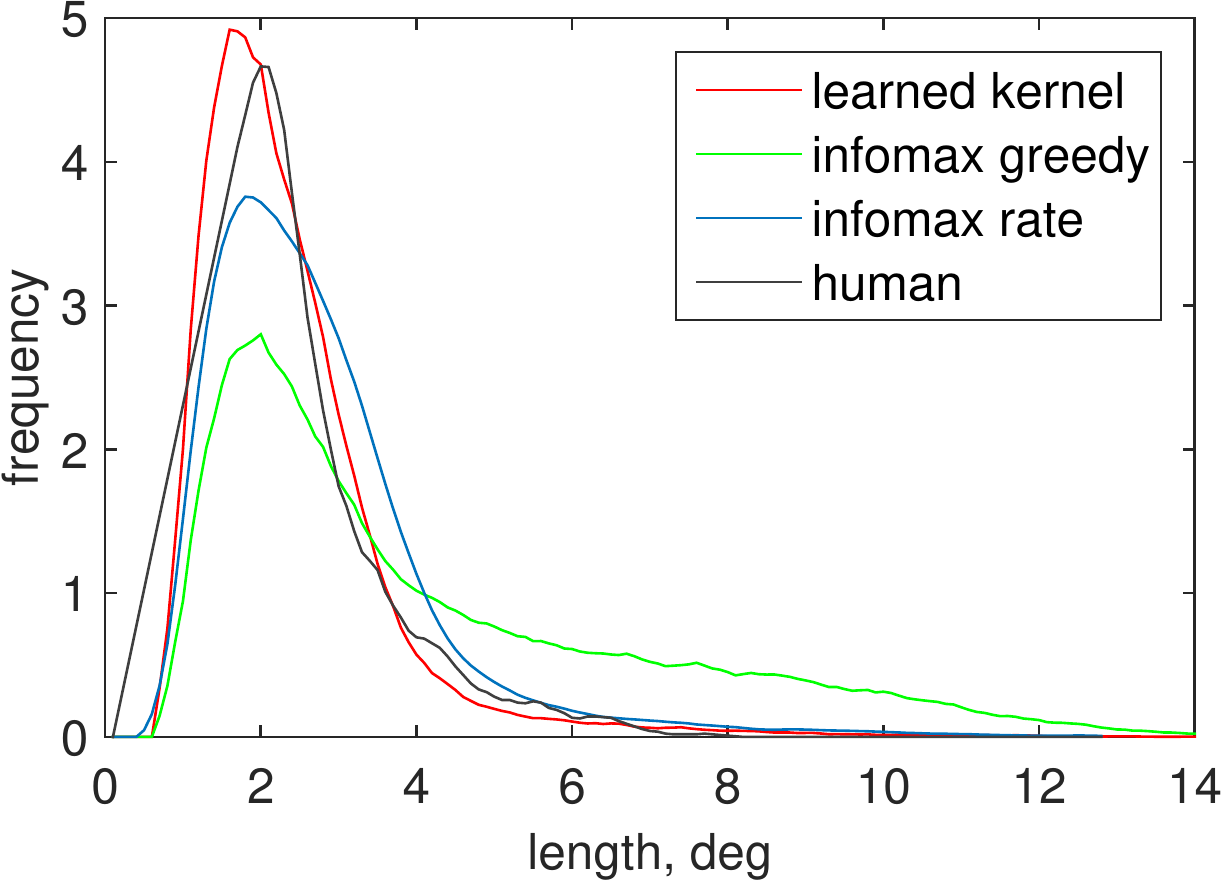}\hfill\includegraphics[scale=0.43]{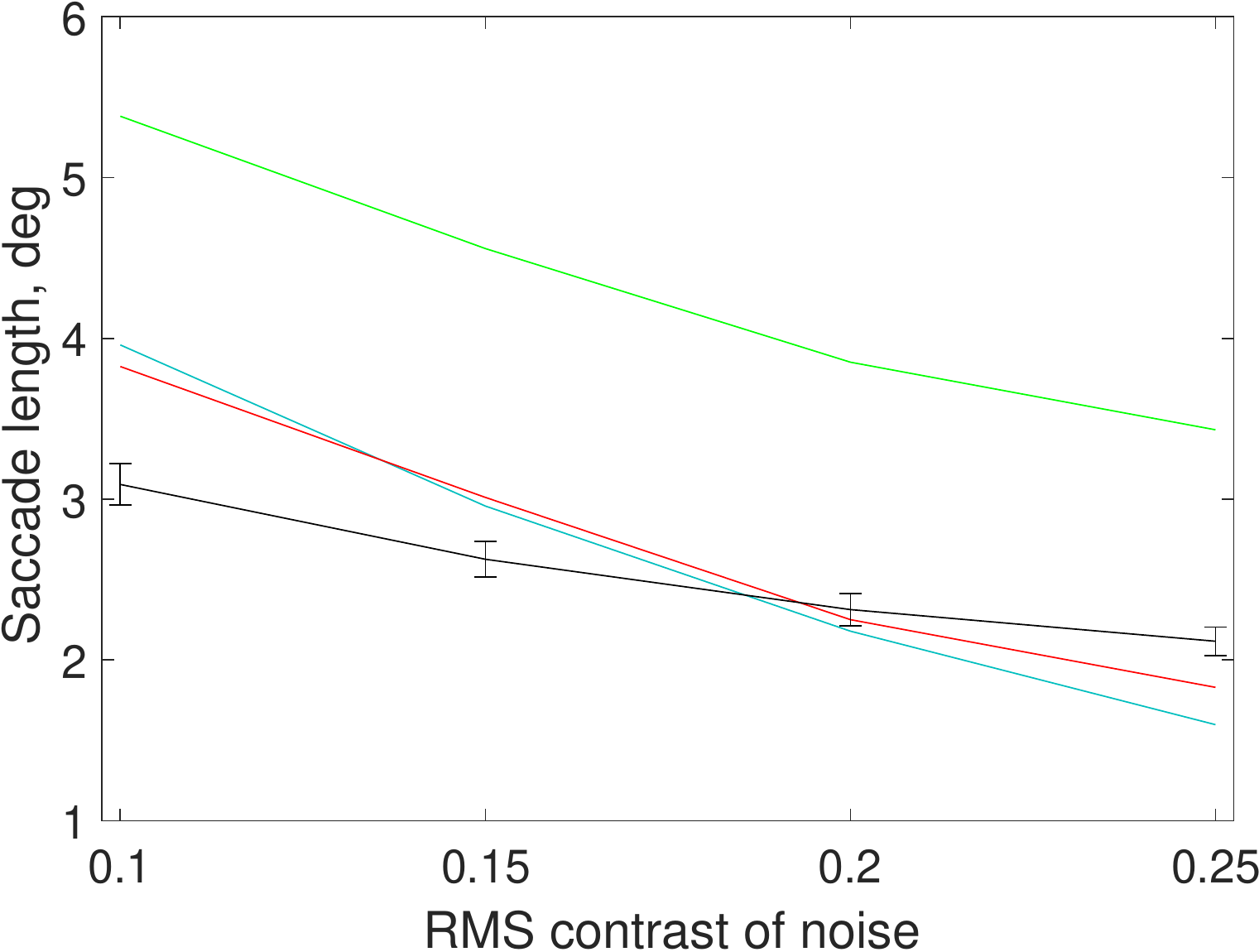}\hspace*{\fill}\caption[The length distribution of saccadic events.]{\label{fig:Length-distribution-of} The histograms of the length distribution
of the saccadic events (left) for trajectories generated under the policies
$\pi_{0},\pi_{1}$ and $\mu_{conv}$ and the human eye-movements corresponding
to the experimental conditions: $e_{n}=0.2$, $e_{t}=0.2$. The data was binned with the resolution of $0.1 \deg$. The distribution
function for all policies and human observer exhibits an ascent between
$0 \deg$ and maximum around $2$ $\mathrm{deg}$. The distribution of
length corresponding tothe  infomax greedy $\pi_{0}$ stabilizes after
$4$ $\mathrm{deg}$ and declines only after $10$ $\mathrm{deg}$.
It is not consistent with length distribution of human saccadic eye-movements,
which is concave on an interval $\left[4.0\protect\textdegree,14.0\protect\textdegree\right]$.
The mean length of saccades decreases with $e_{n}$ (right). It's
immediate consequence of the decrease of the width of FPOC with $e_{n}$,
which defines the area of inhibition from the subsequent fixations.}

\end{figure*}

The Figure \ref{fig:Length-distribution-of} (left) shows the length distributions
of saccades of the human observers and the simulated agents performing the visual search task corresponding to the experimental conditions:  $e_{n}=0.2$, $e_{t}=0.2$.

The distributions for all policies and the human observers exhibit an ascent
between $0\deg$ and maximum around $2\deg$.
The difference in the behaviour of the distributions starts from $4\deg$.
In this experimental conditions the share of the saccades of the human observer
with the length larger than $4\deg$ is $18\%$, whereas this value
for $\pi_{0}$ is $38\%$.
The length distribution for $\pi_{0}$ stabilizes on the interval $[4.0,14.0]\deg$
 that was observed in the earlier work \cite{najemnik2005optimal},
and we found that length of this "stability" interval increases linearly with the grid size.
The reason behind this is an uniform radial ranking of policy $\pi_{0}$
for all locations due to the constant radial function (\ref{eq:pi0}).
The decline of probability starts only at a distance compared to the
size of visual field. 

On the other hand the length distributions of trajectories under $\mu,\pi_{1}$
are concave on an interval $[4.0,14.0]\deg$,
which is also a characteristic for human eye-movement \cite{tatler2006long,castelhano2009viewing}.
The behaviour of the radial function of $\mu$ reflects the non-uniform radial
ranking (a preference in decision-making, see figure (\ref{fig:Where-to-fixate-1}))
of the locations. As a result, the remote locations have significantly lower
probabilities to be chosen as the next destinations.
\par We performed Kolmogorov-Smirnov (K-S) test to check equality of distributions of experimental and simulated saccades. The results of the test are summarized in table  \ref{K-S test} .  The first and the second columns show the values of RMS contrast in the psycho-physical experiment and corresponding critical values of the test statistics at a significance level $\alpha=0.01$. The critical values are different due to the difference in number of saccades for each experimental condition. The next three columns show K-S test statistics for the distributions of the saccades simulated under the different policies. K-S test indicated a higher statistical similarity between the distributions of the experimental saccades and the saccades simulated under the learned policy $\mu$  for the cases $e_{n}=(0.15, 0.2, 0.25)$. In the case $e_{n}=0.1$ the infomax rate and the learned policy explained the experimental distribution equally well.   From this, we can make a conclusion that simulations under the  learned policy explained the best the length distribution of the human eye-movements. 
\begin{table}
\centering
\begin{tabular}{|c | c |c | c | c | } 
\hline 

\multicolumn{1}{|c|}{Cases} & \multicolumn{4}{|c|}{K-S test statistics } \\
\hline
\thead {noise \\ contrast} &\thead{critical \\ value}& \thead{infomax \\ greedy} & \thead{infomax \\ rate} & \thead{learned \\policy} \\
\hline 
0.1 & 0.067 & 0.312 & 0.121&0.124 \\ \hline
0.15  & 0.052 &0.256 & 0.082 &0.081 \\ \hline
0.2 & 0.047 &0.212 & 0.053  & 0.04 \\ \hline
0.25 & 0.035& 0.201 & 0.095  &0.074 \\
[1ex]
\hline 

\end{tabular}
\caption{Statistics for K-S test between the experimental distribution of saccade length and simulated distributions for different policies. The first and the second columns show the values of RMS contrast in the psycho-physical experiment and corresponding critical values of test statistics for significance level $\alpha=0.01$. The next three columns show K-S test statistics for distribution of saccades simulated under different policies.  }
\label{K-S test}
\end{table}
\par The mean length of the saccades was estimated from $10^{4}$ episodes of PO-MDP for all three policies and compared with the mean length of the saccades of  the human observers (see figure
\ref{fig:Length-distribution-of} right). According to our results, the mean length of the saccades decreases
with $e_{n}$, which is consistent with our simulations. It's an immediate consequence
of the decrease of values of FPOC with the increase of the RMS contrast of noise,
which is illustrated on figure \ref{fig:Fovea-Peripheral-Operating-Characteristics}.
The amplitude of the signal exceeds the amplitude of noise within the
circle area with radius $r$ that satisfies the condition $F(r)=1$
(we call this radius the "width of FPOC''). This circle area is effectively
inhibited from the subsequent fixations (see figures (\ref{fig:Where-to-fixate})
and (\ref{fig:Where-to-fixate-1})), because information is already
gathered with a sufficient level of confidence. However, we found that our model provides close estimates of the mean length only for high values of the RMS contrast of noise. Our experimental findings are consistent with previously reported results \cite{geisler2006visual}, where the visual search experiments were set for several levels of the RMS contrast of background noise. In future works we plan to incorporate more complex saccade execution model that takes into account the bias toward the optimal saccade length \cite{engbert2005swift} in order to explain a lower variability of the saccade length in the experiments.  

\subsection{Geometrical persistence }

In this section we analyze the distribution of the directional angle
$\theta_{d}$ (this notation was introduced in \cite{amor2016persistence})
of the human saccadic eye-movements and the simulated trajectories. The directional
angle is the angle between two consequent saccades, and, therefore,
can be defined as $\theta_{d}=\tan^{-1}\left(y_{n+1}/x_{n+1}\right)-\tan^{-1}\left(y_{n}/x_{n}\right)$,
where $\left(y_{n},x_{n}\right)$ are the  coordinates of $n$th fixation.
According to this definition, the movement is related to a persistent
one if the directional angle is close to $0$ or $2\pi$. The angles with
the values close to $\pi$ correspond to anti-persistent movements. 

\begin{figure*}
\hspace*{\fill}\includegraphics[scale=0.47]{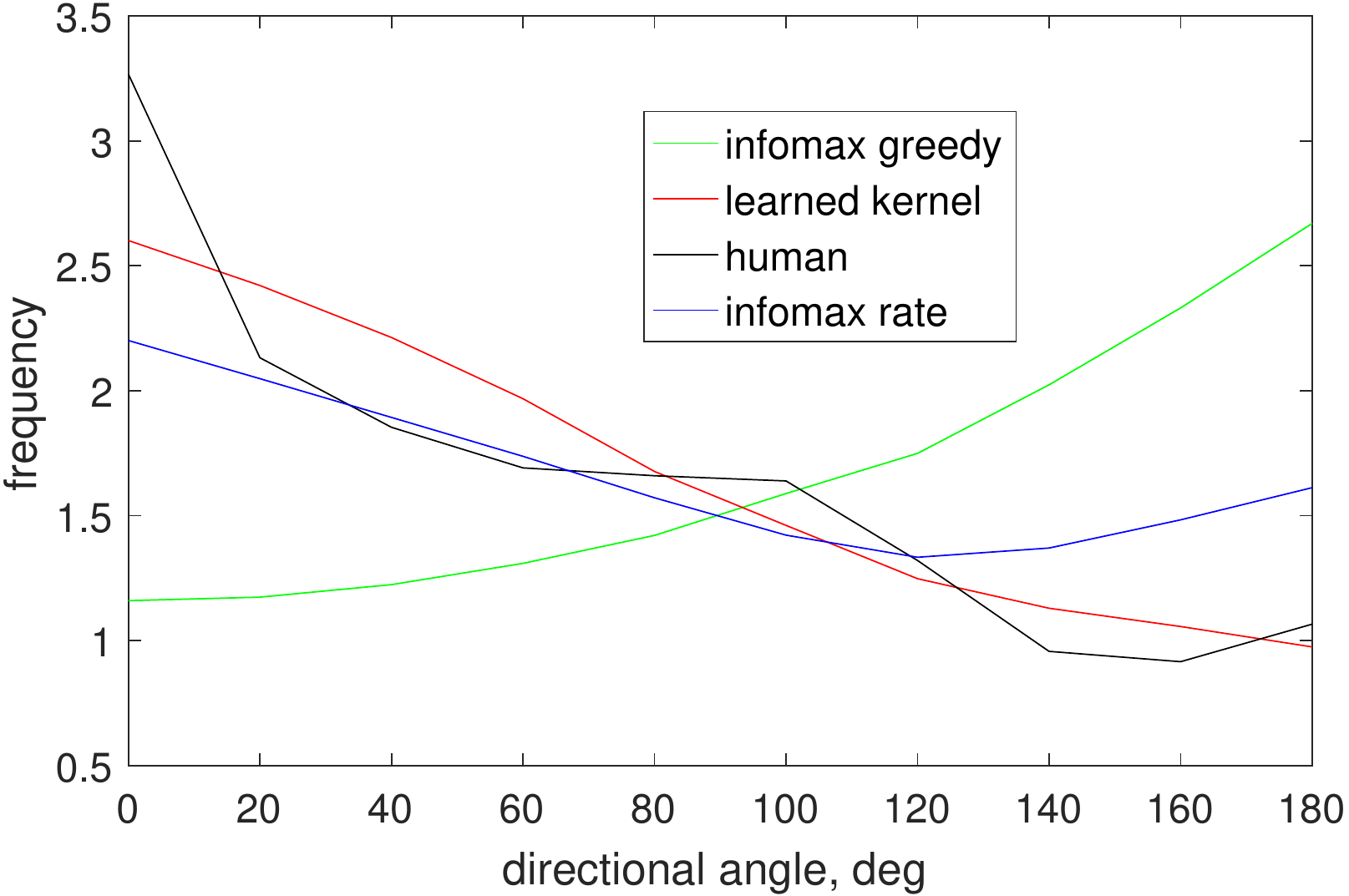}\hfill\includegraphics[scale=0.52]{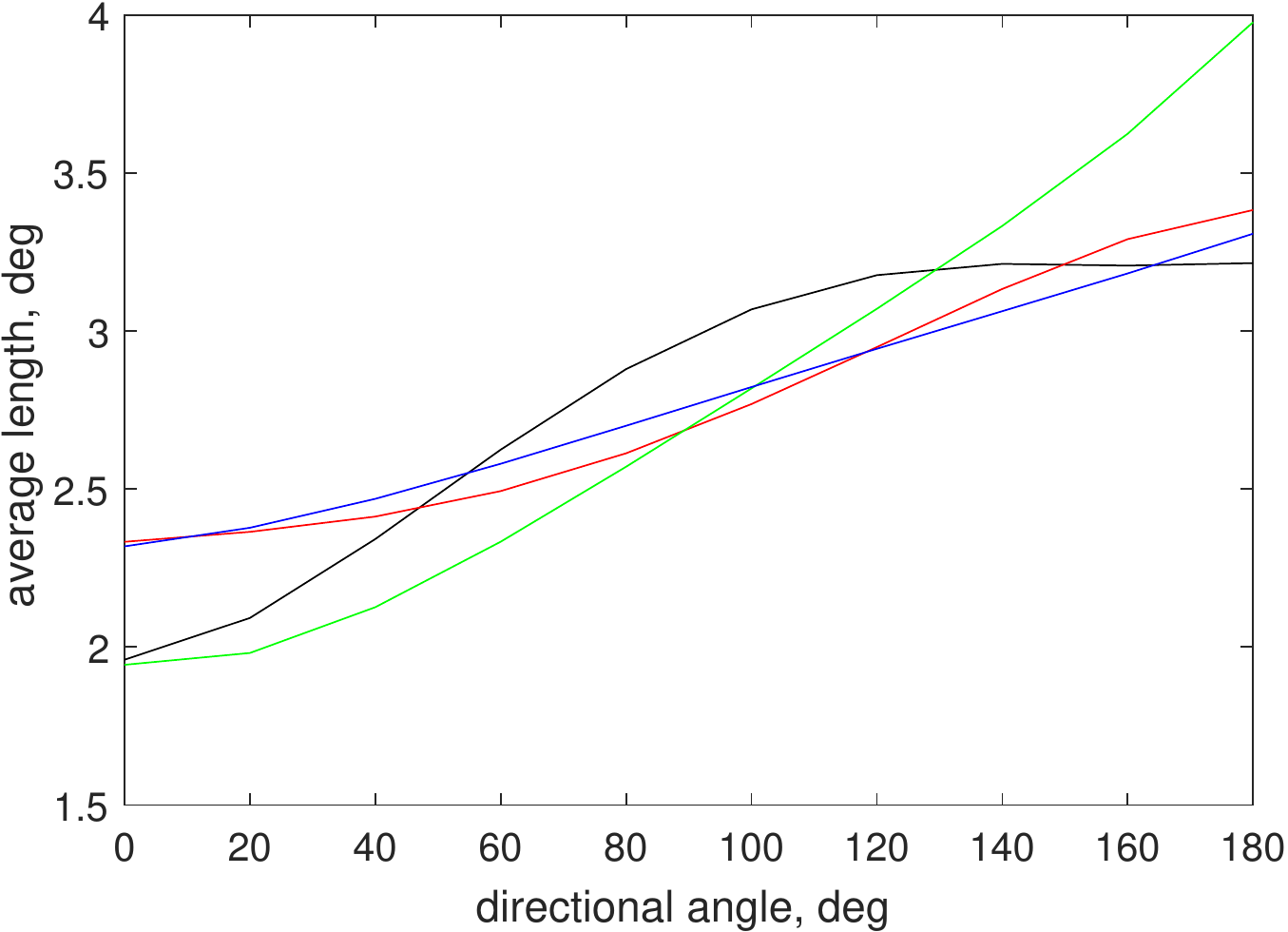}\hspace*{\fill}\caption{\label{fig:Directional-angle-distribution} The histograms of the directional angle
(left) and the distributions of the mean length of the saccades to the directional
angle (right). Data for both histograms was binned with the resolution of 20 $\deg$. The infomax greedy policy $\pi_{0}$ generates the
trajectories with stable anti-persistent movements (left), with a high
degree of separation between large and small movements (right). In
contrast, the decision-making process under the infomax rate policy $\pi_{1}$
tends to preserve the direction of movement. The dynamics under the learned policy $\mu$ is also characterized as a persistent random
walk. }
\end{figure*}

The distributions of the directional angle were calculated for the trajectories
generated by Markov decision process with the policies $\pi_{0},\pi_{1}$
and $\mu$. Figure (\ref{fig:Directional-angle-distribution})(left)
demonstrates the distribution of the directional angle of the saccadic events for the
human observers and the simulated trajectories for $e_{n}=0.2$ and $e_{t}=0.2$.
The infomax greedy policy $\pi_{0}$ generates the trajectories with
stable anti-persistent movements, because the policy $\pi_{0}$ chooses the
next fixation location without taking the current location into consideration.
Due to the inhibitory behavior of infomax, it's much less likely to choose
the nearby location instead of remote and relatively unexplored ones.
Only geometrical borders limit the choice of the next fixation, which
results in fixations on the opposite side of the visual field (as the most
remote point, look at the figure (\ref{fig:Where-to-fixate})). 

In contrast, the decision process under the learned policy $\mu$ tends
to preserve the direction of the movement. The dynamic of the system under the
policy $\mu$ is quite similar to self-avoiding random walk model
described in \cite{engbert2011integrated}. Due to the asymptotic
behavior of the kernel function $K(x,y)$, the reward gain from the remote locations
is suppressed, meanwhile, the locations, which are already visited,
are also inhibited (look at the figure (\ref{fig:Where-to-fixate-1})).
This results in short-range self-avoiding movements, which demonstrate
the persistent behavior \cite{isogami1992structural,engbert2011integrated},
and, therefore, the probability distribution of the directional angle $\theta_{d}$ is biased
towards values $0$ or $2\pi$. According to the Figure \ref{fig:Directional-angle-distribution}
(left), the dynamics under the heuristics $\pi_{1}$ is also characterized
as a persistent random walk. The learned policy $\mu$ has, in general,
a stronger radial ranking of locations than $\pi_{1}$, which results
in a shorter range of saccades, and a repulsion, caused by inhibition,
becomes more relevant. The distribution of average length of saccades
depending on $\theta_{d}$ is shown on Figure \ref{fig:Directional-angle-distribution}
(right). On average the co-directed movements are shorter than the reversal
ones for all policies.

\begin{figure}
\centering
\includegraphics[scale=0.55]{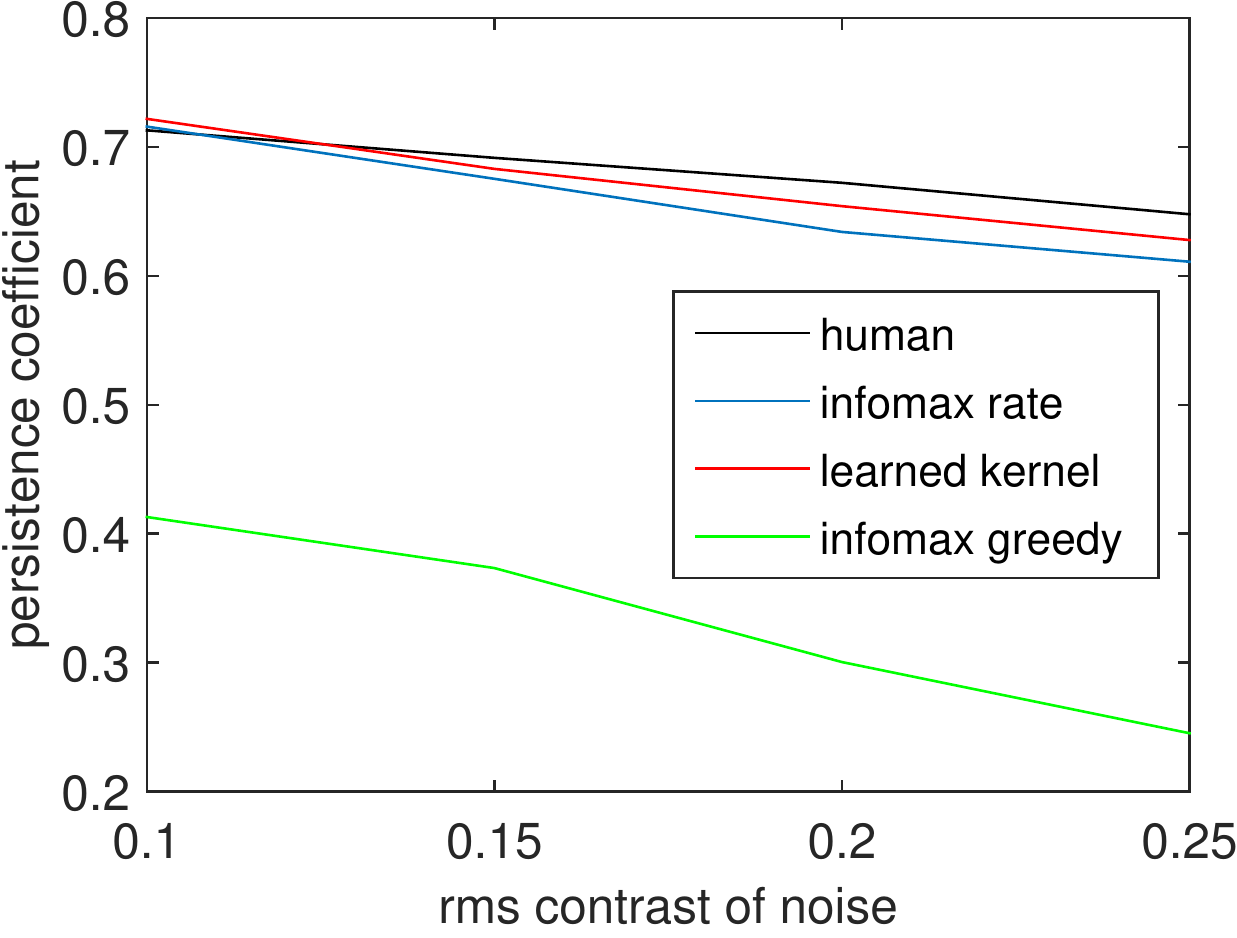}\caption{\label{fig:Dependence-of-persistence}The share of saccades, which
retain the direction of the previous movement: $cos(\theta_{d})>0$, is
called the "persistence coefficient". This quantity demonstrates the dependence of the persistence 
on the visibility of the target. As it was mentioned previously, the average
saccade length is decreasing with the growth of the RMS contrast (\ref{fig:Length-distribution-of}).
Therefore, the linear term (\ref{eq:duration_fix}) in the duration of the
steps becomes less relevant, and the decision-making becomes more agnostic
about temporal costs (closer to the information greedy $\pi_{0}$). The
decline of the persistent coefficient is also a characteristic of the human
eye movements, which was not covered in the previous research. }

\end{figure}

In our experiments we discovered that the geometrical persistence depends on the visibility of target (on FPOC in the simulations).
We measured the share of the saccades, which retain the direction of the previous
movement: $\cos(\theta_{d})>0$. This quantity is called ``persistence
coefficient''. The figure \ref{fig:Dependence-of-persistence} demonstrates
the dependency of the persistence coefficient on the RMS contrast of background
noise for the human observers and the simulated trajectories. As it was mentioned
previously, the average saccade length is decreasing with the growth of
RMS contrast (\ref{fig:Length-distribution-of}). Therefore, the linear
term (\ref{eq:duration_fix}) in the duration of steps becomes less relevant,
and the decision-making becomes more agnostic about  the temporal costs (closer
to the information greedy $\pi_{0}$). The decline of the persistent coefficient
is also a characteristic of human eye movements, which was not covered
in the previous research. 

\section{Statistical persistence}

In the previous section we have analyzed the geometrical persistence of
the human eye-movements and the trajectories simulated under three different policies. However, 
this statistical property doesn't give any insight into a long-range
correlation in time-series. In this section we show that dynamics
under the learned policy $\mu$ have a multifractal behavior,
which is similar to that of the human eye-movements during execution of the visual search task
.\par In contrast to the previous research \cite{amor2016persistence} in our analysis we distinguish between two different types of the multifractality by a calculation of a generalized Hurst exponent for shuffled time series. We separate the time series on fixational and saccadic eye-movements, which allows us to demonstrate the fundamental difference in the temporal structure of these types of eye-movements.  It was shown that the behaviour of the generalized Hurst exponent is consistent with the basic statistical properties of eye-movements. After this we demonstrate that the dynamics under the optimal policy of gaze allocation explains the changes in scaling behaviour of eye-movements with difficulty of the visual task both on qualitative and quantitative levels.
\par For statistical analysis of simulated
trajectories we use a multifractal detrended fluctuation analysis (MF-DFA)\cite{kantelhardt2002multifractal},
which is a widely-used method for detection of long-range correlations
in stochastic time-series. It has found successful applications in
the field of bioinformatics \cite{rosas2002multifractal,dutta2013multifractal},
nano and geo-physics \cite{vandewalle1999non}. This method is based
on the approximation of trends in time-series and the subtraction of detected
trends (detrending) from original data on different scales. The detrending
allows deducting the undesired contribution to long-range correlation,
which is a result of non-stationarities of physical processes. We use
the package provided by Espen Ihlen \cite{ihlen2012introduction}
for all our estimations of the generalized Hurst exponent in this section. 
\par In the appendix \ref{sec:Multifractal-analysis} we thoroughly explain the details of the multifractal analysis. The subsection \ref{sub:Multifractal-analysis-intro} presents the details of MF-DFA algorithm.  In the subsections \ref{sub:mf_simulated}  and \ref{sub:mf_simulated} we explain how MF-DFA is performed over the simulated trajectories.   The results of the multifractal analysis of the human eye-movements are presented in \ref{sub:Multifractality-of-human}. The subsection \ref{sub:dependence_vis} summarizes our findings and compares the  generalized Hurst exponent of the simulated trajectories to one of human eye-movements for different experimental conditions. 

\subsection{Multifractality of human eye movements\label{sub:Multifractality-of-human}}

We perform MF-DFA over the difference of time series of the human gaze positions
and in order to compare the estimated generalized Hurst exponent with the
simulations. The differentiated time series was estimated from raw
data of coordinates of the gaze fixations $A=\left\{ \left(x_{1},y_{1}\right),...\left(x_{N},y_{N}\right)\right\} $with the
resolution of $7$ ms:

\begin{equation}
\Delta X=\left\{ \left(x_{2}-x_{1}\right),\ldots\left(x_{N}-x_{N-1}\right)\right\} 
\end{equation}

\begin{equation}
\Delta Y=\left\{ \left(y_{2}-y_{1}\right),\ldots\left(y_{N}-y_{N-1}\right)\right\} 
\end{equation}

The time series $\Delta X$ and $\Delta Y$ were estimated for
each trial with certain experimental conditions and concatenated over
all participants. After this, we represent the differentiated time
series in the following way: $\Delta X=\left\{ F_{1},S_{1},....,F_{m-1},S_{m-1},F_{m}\right\} $,
where $F_{i}$ and $S_{i}$ correspond to the sequences of the movements during
time interval of $i$-th fixation and saccade respectively \cite{amor2016persistence}.
We separate the differentiated time series on the fixational and the saccadic
time series:

\begin{equation}
\Delta X_{F}=\left\{ F_{1},0_{s_{1}},\ldots F_{m-1},0_{s_{m}},F_{m}\right\} 
\end{equation}

\begin{equation}
\Delta X_{S}=\left\{ 0_{f_{1}},S_{1},\ldots 0_{f_{m-1}},S_{m},0_{f_{m}}\right\} 
\end{equation}
where  $0_{n}$ corresponds to zero array with the length $n$, and $f_{m}$ and $s_{m}$ are the lengths of corresponding sequences $F_{m}$ and $S_{m}$. 

\begin{figure*}
\centering
\includegraphics[scale=0.77]{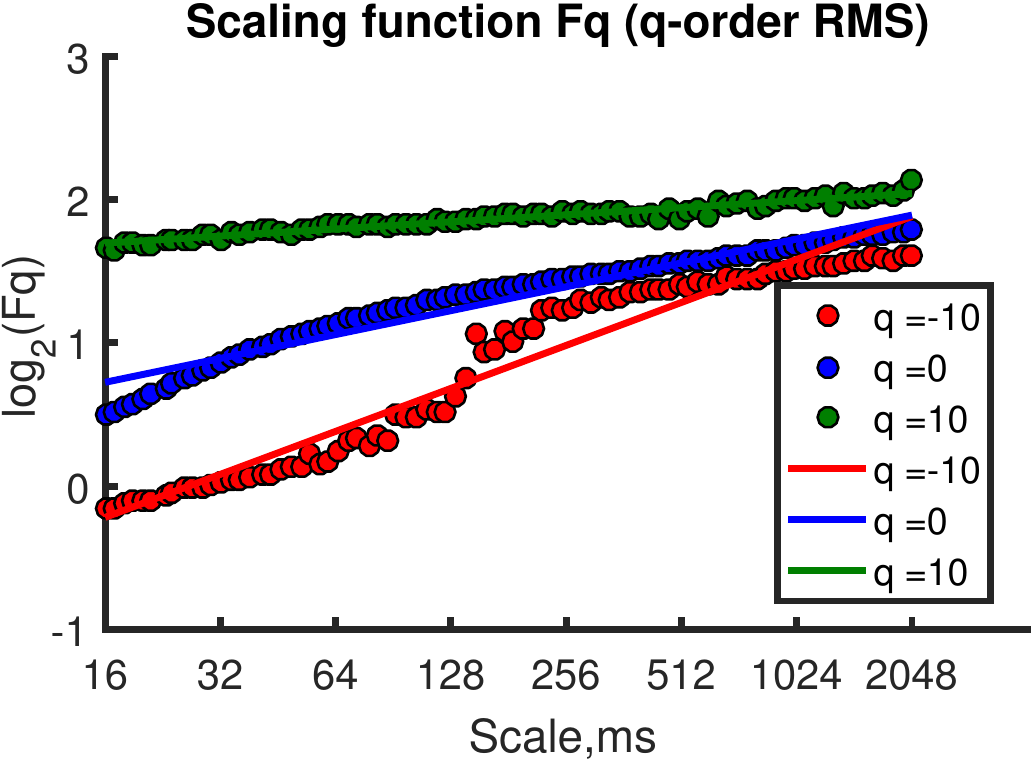}\includegraphics[scale=0.77]{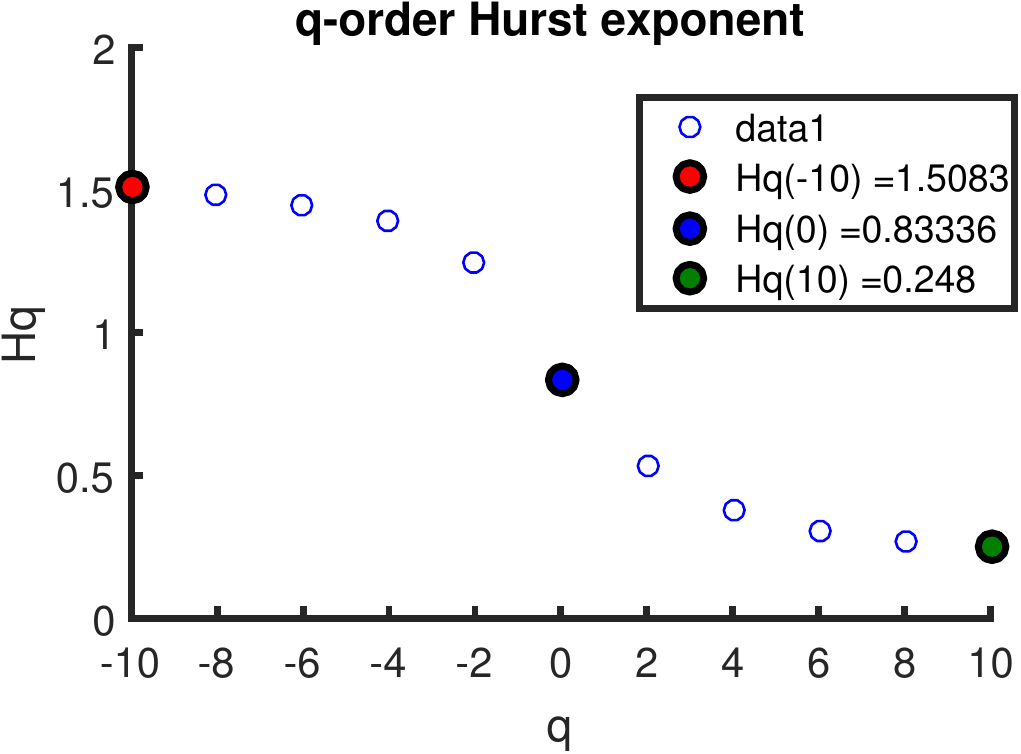}\caption{\label{fig:crossover}The scaling of the q-order fluctuation function
$F_{q}(s)$ (left), and the generalized Hurst exponent $H(q)$ (right)
computed through a linear regression of $\log_{2}\left(F_{q}(s)\right)$.
This graph is a result of application of MF-DFA over the horizontal differentiated
time series $\Delta X$ of the concatenated human scan-paths for the experimental
conditions: $e_{t}=0.2$, $e_{n}=0.25$. 
The red, blue and green lines correspond to the linear approximation of
function $\log_{2}\left(F_{q}(s)\right)$ for the orders $q=\left\{ -10;0;10\right\} $.
The scaling of $F_{q}(s)$ exhibits the crossover the crossover on a time scale  of $256$ $ms$. The crossover separates the "lower"
and the "upper" regimes mentioned in \cite{amor2016persistence}.
The lower regime is related to fixational eye-movements (which is
supported by the value of crossover scale s$_{cros}$ being close to average
fixation duration), and the upper regime - to saccadic ones. The crossover
in the scaling of $F_{q}(s)$ was observed for all experimental conditions. }

\end{figure*}

The figure \ref{fig:crossover} demonstrates the scaling of the q-order fluctuation
function $F_{q}(s)$ (\ref{eq:fluctuation_function}). This graph
is a result of the application of MF-DFA over the horizontal concatenated
differentiated time series $\Delta X$ of the human scan-paths for the experimental
conditions: $e_{t}=0.2$, $e_{n}=0.25$. 
The red, blue and green lines correspond to the linear approximation of
function $\log_{2}\left(F_{q}(s)\right)$ for the orders $q=\left\{ -10;0;10\right\} $.
The scaling of $F_{q}(s)$ exhibits the crossover on a time scale  of $256$ $ms$. The crossover separates the "lower''
and "upper'' regimes mentioned in \cite{amor2016persistence}.
According to Amor et. al. the crossover is caused by the presence of two
different generative mechanisms of eye-movements. The lower regime
is related to fixational eye-movements (which is supported by the value
of crossover scale s$_{cros}$ being close to the average fixation duration),
and upper regime - to the saccadic ones. The crossover in the scaling of $F_{q}(s)$
was observed for all experimental conditions. The value of generalized
Hurst exponent $H(q)$ (Figure \ref{fig:crossover} right) is obtained
through linear regression of $\log_{2}\left(F_{q}(s)\right)$. Our
estimates of $H(q)$ are consistent with the ones of Amor et. al.
for both directions and all regimes. 

In order to distinguish between two different types of  multifractality
\cite{kantelhardt2002multifractal} we calculated the generalized Hurst exponent $H_{shuf}(q)$ for
the shuffled differentiated time series. The first type of multifractality
is a consequence of a broad probability density function for the values
of time series. If only multifractality of the first type presents in
time series: $H(q)=H_{shuf}(q)$. The second type of  multifractality
is caused by the difference in correlation between large and small fluctuations,
which is a scenario described in \cite{amor2016persistence}. In this
case $H_{shuf}(q)=0.5$ and $H(q)=0.5+H_{corr}(q)$, where $H_{corr}(q)$
is (negative) positive for the long-range (anti-)correlation. If both
types of multifractality present in time series: $H(q)=H_{shuf}(q)+H_{corr}(q)$. 

\begin{figure*}

\centering
\includegraphics[scale=0.87]{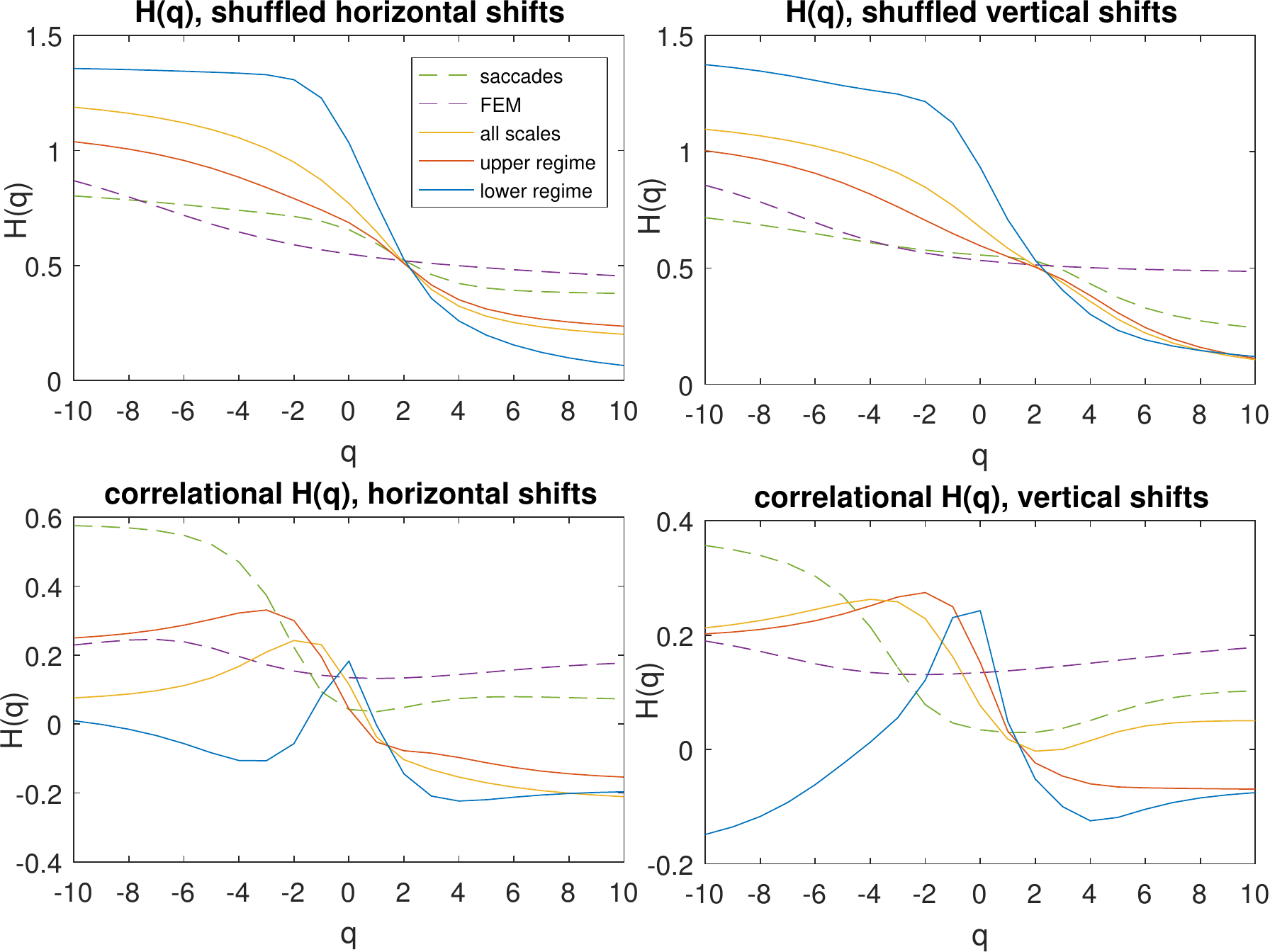}

\caption{\label{fig:Correlational-generalized-Hurst}The Hurst exponent of the shuffled
time series $H_{shuf}(q)$ (top) and the correlational Hurst exponent
$H_{corr}(q)$ (bottom) for the horizontal (left) and the vertical components
(right) of the human eye-movements. As well as a previous graph \ref{fig:crossover},
this one is a result of an application of MF-DFA over the concatenated human
scan-paths for the experimental conditions: $e_{t}=0.2$, $e_{n}=0.25$.
The behaviour of $H_{shuf}(q)$ for both horizontal and vertical shifts
for full scales corresponds to the one mentioned in \cite{kantelhardt2002multifractal}
(eq. 27). We assume that multifractality of the first type is caused by an
asymptotic behaviour of the amplitude distribution of saccades (see
figure \ref{fig:Length-distribution-of}). The difference in the long-range
correlation of large and small fluctuations is reflected by $H_{corr}(q)$
(figure \ref{fig:Correlational-generalized-Hurst} bottom). Due to
the properties of the fluctuation function \ref{eq:fluctuation_function}
for positive (negative) $q$-orders the main contribution are coming
from segments containing small (large) fluctuations \cite{kantelhardt2002multifractal}.
The positive (negative) long range correlation ($H_{corr}(q)>0$)
is, therefore, a characteristic of small (large) fluctuations in the upper
and the full scales regimes for both directions. In general, these results
are consistent with the distribution of the average length of saccades to the directional
angle (see figure \ref{fig:Directional-angle-distribution} right),
which also indicates the difference in persistence of large and small
saccades. }
\end{figure*}

The figure \ref{fig:Correlational-generalized-Hurst} demonstrates our
estimates of the Hurst exponent of the shuffled time series $H_{shuf}(q)$
(top) and the correlational Hurst exponent $H_{corr}(q)$ (bottom) for
the horizontal (left) and the vertical components (right). We estimated both
exponents for the saccades (green dashed line) and FEM (purple dashed
line) in the upper and the lower regimes of scales respectively. As well as
a previous graph \ref{fig:crossover}, this one is a result of an application
of MF-DFA over the concatenated differentiated time series of the human eye-movements
for the experimental conditions: $e_{t}=0.2$, $e_{n}=0.25$. The behaviour
of $H_{shuf}(q)$ for the full  time series and the saccadic time series in the upper regime
corresponds to the one mentioned in \cite{kantelhardt2002multifractal}
(eq. 27):

\begin{equation}
H(q)\sim\begin{cases}
\begin{array}{c}
1/q\,\,\,\,\left(q>\alpha\right)\\
1/\alpha\,\,\,\,\left(q\leq\alpha\right)
\end{array}\end{cases}\label{eq:Levy}
\end{equation}
with $\alpha\sim1$. The equation $\ref{eq:Levy}$ was derived for
time series of uncorrelated random values with the power law distribution: 

\begin{equation}
P=\left\{ \begin{array}{c}
\alpha x^{-(\alpha+1)}\,\,\,\,x\geq1\\
0\,\,\,\,\,x<1
\end{array}\right.\label{eq:power}
\end{equation}

One can see a similarity of the function (\ref{eq:power}) with the distribution of the amplitude of the saccadic events for humans (see
figure \ref{fig:Length-distribution-of}). The amplitude distribution
of the saccades demonstrates the power law behavior on the interval $\left[4.0\textdegree,14.0\textdegree\right]$
with $\alpha\approx1$. 
The probability distribution function (\ref{eq:power}) also reflects
an absence of saccades with the length lower than minimal one. Therefore,
the first type of multifractality of the saccadic time series is caused by the broad probability
distribution of saccade magnitude. 

The difference in the long-range correlation of large and small fluctuations
is reflected by $H_{corr}(q)$ (figure \ref{fig:Correlational-generalized-Hurst}
bottom). Due to the properties of fluctuation function (\ref{eq:fluctuation_function})
for the positive (negative) $q$-orders the main contribution are coming
from segments containing the large (small) fluctuations \cite{kantelhardt2002multifractal}.
The positive (negative) long-range correlation ($H_{corr}(q)>0$)
is, therefore, a characteristic of the small (large) fluctuations in the upper
regime for the saccadic and the full time series. These results
are consistent with the distribution of the average length of saccade to the directional
angle (see figure \ref{fig:Directional-angle-distribution} right),
which also indicates the difference in the persistence of large and small
saccades. Therefore, we confirm here that the small saccadic eye-movements
demonstrate the long-range correlations as well as fixational eye-movements. 

The time series of FEM demonstrates the monofractal behaviour and the positive
correlations with $H\approx0.8$ in the lower regime of scales \cite{amor2016persistence}.
However, the behaviour of both $H_{corr}(q)$ and $H_{shuf}(q)$ for
the full time series in the lower regime indicates the presence of multifractalities
of both types. At the present moment we have no explanation of the
multifractality in the lower regime and leave this problem for a future
work.

\subsection{Dependence on visibility\label{sub:dependence_vis}}

In this section we present a comparison of the generalized Hurst exponent
for the human eye-movements in the upper regime and the simulated trajectories
under the learned policy. As well as in the case of the geometrical persistence,
we claim the quantitative properties of the statistical persistence depend
on the visibility of the target. 

\begin{figure*}

\includegraphics[scale=0.46]{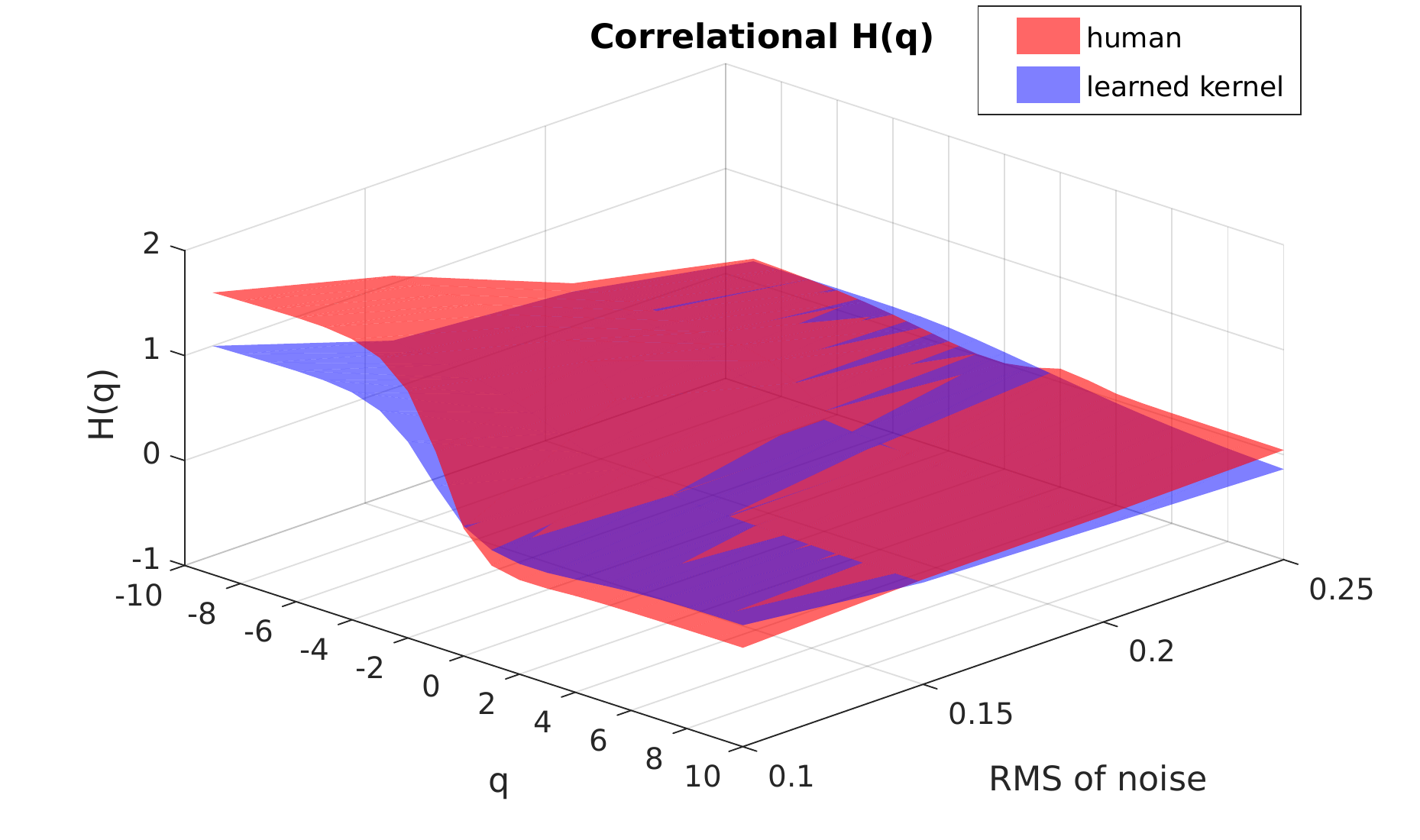}\includegraphics[scale=0.46]{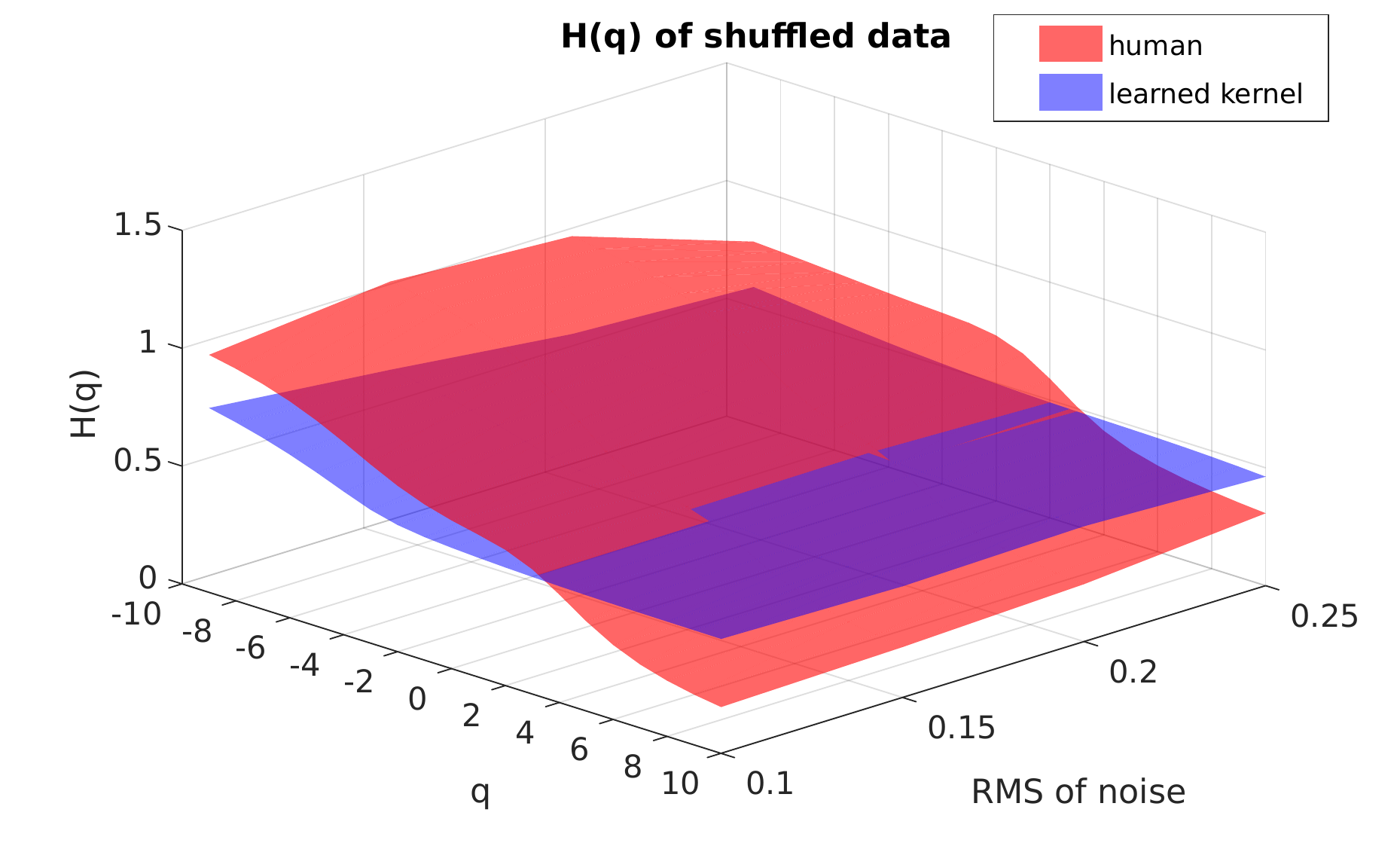}\caption{\label{fig:Dependence-of-Hurst}This figure demonstrates $H_{corr}(q)$
(left) of the simulated trajectories (blue) under the learned policy $\mu$
and the correlational Hurst exponent for the human eye-movements (pink) averaged
over two directions: $H_{corr}(q)=\left(H_{corr}^{x}(q)+H_{corr}^{y}(q)\right)/2$
in the upper regime. The correlational Hurst exponents for negative $q$-orders
declines with the growth of RMS contrast of noise both for human eye-movements
and the simulated trajectories.  For the positive $q$-orders the correlational
Hurst exponent is less affected by the change of the visibility of the target.
In general, the correlations weaken with the growth of the RMS contrast, which
is consistent with the decline of the geometrical persistence \ref{fig:Dependence-of-persistence}.
The Hurst exponent of the shuffled time series (right), as well as the correlational
Hurst exponent, demonstrates the decline with the growth of the RMS contrast for
negative q-orders both for the human eye-movements and the simulated trajectories.
In the subsection (\ref{sub:Multifractality-of-human}) we mentioned that
the behaviour of $H_{shuf}(q)$ resembles the one related to time series
of random values with power law distribution \ref{eq:power}. }
\end{figure*}

We estimated the correlational Hurst exponent $H_{corr}(q)$ and the Hurst
exponent of the shuffled time series $H_{shuf}(q)$ for the differentiated
trajectories of the human eye-movements for all levels of the RMS contrast
of background noise: $e_{n}\in\left(0.1,0.15,0.2,0.25\right)$. Figure
\ref{fig:Dependence-of-Hurst} (left) shows $H_{corr}(q)$ (left)
of simulated trajectories (blue) under the learned policy $\mu$ and the correlational
Hurst exponent for the human eye-movements (pink) averaged over two directions:
$H_{corr}(q)=\left(H_{corr}^{x}(q)+H_{corr}^{y}(q)\right)/2$ in the upper
regime. The correlational Hurst exponents for the negative $q$-orders
declines with the growth of the RMS contrast of background noise both for the human eye-movements
and the simulated trajectories. This indicates the weakening of the correlation
between small fluctuations. For the positive $q$-orders the correlational
Hurst exponent is less affected by the change of the visibility of target.
The $H_{corr}(q)$ for $q=10$ stabilized on values $0.04$ and $-0.12$
for human eye-movements and the simulated trajectories correspondingly.
In general, the correlations weaken with the growth of the RMS contrast, which
is consistent with the decline of the geometrical persistence \ref{fig:Dependence-of-persistence}.
The decline of the Hurst exponent with the increase of difficulty of visual
search task was also observed in the previous work \cite{stephen2011fractal}. 

The Hurst exponent of the shuffled time series (Figure \ref{fig:Dependence-of-Hurst}
right), as well as the correlational Hurst exponent, demonstrates the decline
with the growth of the RMS contrast for the negative q-orders both for the human eye-movements
and the simulated trajectories. In the subsection \ref{sub:Multifractality-of-human}
we mentioned that the behaviour of $H_{shuf}(q)$ resembles the one related
to time series of random values with the power law distribution \ref{eq:power}.
The average value of this time series equals $1/\left(\alpha-1\right)$
for $\alpha>1$. The increase of $\alpha$ results both in the decrease
of the average value in time series and the decrease of the value of $H_{shuf}(q)\sim1/\alpha$
for $q<0$. Therefore, the average value in time series and the values
of $H_{shuf}(q)$ for the negative $q$-orders are correlated in the assumption
of the power-law distribution. Previously we found the decrease of the average
saccade length with the growth of RMS of background noise \ref{fig:Length-distribution-of},
which is consistent with the decrease of values of $H_{shuf}(q)$ for
negative $q$-orders. We assume that this correlation is caused by the
power-law asymptotic behaviour of the length distribution of human
eye-movements (\ref{eq:power}).

\section{Conclusion}

We have presented a computational model of the ideal observer that both
qualitatively and quantitatively describes the human visual behaviour
during the execution of the visual search task. The basis of this model is the observer's
representation of the constraints of its own visual and oculomotor systems.
We demonstrated that a consideration of the temporal costs and uncertainty
of the execution of saccades results in the dramatic change of the basic statistical
properties and the scaling behavior of the simulated time series. 

We performed the multifractal analysis of our data and discovered the
presence of two types of multifractality both in time series of the human
eye-movements and the model simulations. The multifractality caused by the
broad amplitude distribution of the saccades (the first type of multifractality)
makes a significant contribution to the multifractal behaviour of time series,
which was not covered in the previous work \cite{amor2016persistence}.
After the estimation of the correlational part of the Hurst exponent \cite{kantelhardt2002multifractal}
we confirmed the presence of the long-range positive correlations of the small
saccades in the upper regime. On the contrary, the large saccades exhibit the
weak long-range anti-correlations for the model simulations and the human
eye-movements in the upper regime. As well as in the case of the geometrical
persistence, we found that the long-range correlations between eye-movements
weaken with the decline of the target's visibility, which is consistent with
the previous work on this topic \cite{stephen2011fractal}. 
\par In this research we focused our attention more on the persistence of eye-movements
rather than on their spatial distribution. That's why we didn't consider
the factors that are not directly related to the trade-off between
the temporal costs and the expected information gain. We estimate the optimal
policy under the assumption that the visual search process is characterized
by shift-rotational symmetry \cite{butko2008pomdp}, which was not
observed in the previous work with similar experimental settings \cite{najemnik2008eye}.
The symmetry of the visual search can be broken by angular dependency
of FPOC in both cases of normal controls and patients with vision
disabilities \cite{van2013macular}. We plan to include the angular
dependency to radial and smoothing functions of policy (see eq. (\ref{eq:separability}))
in order to consider the asymmetry of the visual field in our future works.

To sum up, this framework provides an elegant explanation of scaling
and persistent dynamic of the voluntary saccades from an optimality point
of view. It clearly demonstrates that control models are able to
describe human eye-movements far beyond their basic statistical properties.

\appendix

\section{Implementation of reinforcement learning algorithms\label{sec:Implementation-of-reinforcement}}
\subsection{Kernel function} \label{sub:separability}
We assume that the process of the visual search is characterized
by shift-rotational invariance \cite{butko2010infomax}. In this research we focus our attention
on the persistence of eye-movements rather than on
their spatial distribution. That's why we use the approximation of the shift-rotational invariance in which we don't need to
consider the factors that are not directly related
to the trade off between the temporal costs and the expected
information gain, such as an asymmetry of FPOC. 
\par The coefficients in the set of dynamic equations (\ref{eq:observation},\ref{eq:saccade-1},\ref{eq:duration})
are unaltered under any distance preserving transformations. The last
dynamic equation, which is the policy of gaze allocation (\ref{eq:soft_max}),
should be shift-rotational invariant as well. The policy (\ref{eq:soft_max})
is determined by function of an expected reward $f(D,p)$. Due to the property of shift invariance we can represent the function of an expected reward with Volterra series \cite{volterra1944theory}:
\begin{equation}
f(D,p)=f_{0}+\sum_{n=1}^{N} \sum_{l_{1}=1}^{L} \cdots \sum_{l_{n}=1}^{L} K_{n}(D-l_{1},..,D-l_{n})\prod_{j=1}^{n}p(l_{j})
\label{eq:volterra}\end{equation}
Where $K_{n}(l_{1},..,l_{n})$ are called Volterra kernels. The constant $f_{0}$  is eliminated in the equation  \ref{eq:soft_max}, and, therefore, will not be considered. The dimensionality of Volterra kernels $K_{n}$ scales with the number of the potential locations as $L^{n}$.  The estimation of Volterra kernel for $n	\geq 2$ is computationally unfeasible for the  grid size in our simulations: $L=2^{7} \times 2^{7}$. For this reason we consider only the linear term:
\begin{equation}
f(D,p)= \sum_{l} K(D-l)p(l)
\end{equation}
We do not expect that the estimation of higher order terms will result in improvement of the performance of the policy. The current observation model \ref{eq:observation} is based on independent inputs $W_{l}$ on each individual location, which results in independence of the values of the probability distribution $p_{l}$ for a sufficiently large grid size. Therefore, higher order terms don't provide additional information on the location of the target. 
\par The function of the expected reward
should be computed taking into account the current location of the gaze $A$. Considering its rotational invariance the most general form of this function is: $f(D,p)=\underset{l}{\sum}K(\left\Vert D-l\right\Vert ,\left\Vert D-A\right\Vert )p(l)$.
The softmax policy (\ref{eq:soft_max}) for the function of the expected
reward is:

\[\mu(D_{n},p_{n})\propto\]

\begin{equation}
\exp\left(\underset{l}{\sum}p_{n}(l)K\left(\left\Vert D_{n}-l\right\Vert ,\left\Vert D_{n}-A_{n}\right\Vert \right)\right)\label{eq:generalform}
\end{equation}
 Together with the set of
the equations (\ref{eq:observation},\ref{eq:saccade-1},\ref{eq:duration}),
this form of the policy keeps the evolution of the system invariant under
any distance-preserving transformation. The convolution of the probability
distribution with the kernel function $K(x,y)$ in general form \ref{eq:generalform}
is difficult to optimize, and the problem can be effectively solved
only in a separable approximation:
\begin{align*}
K(\left\Vert D_{n}-A_{n}\right\Vert ,\left\Vert D_{n}-x\right\Vert )
\end{align*}
\begin{equation}
\approx R\left(\left\Vert D_{n}-A_{n}\right\Vert \right)S\left(\left\Vert D_{n}-x\right\Vert \right)\label{eq:separability}
\end{equation}

We call $R$ and $S$ the radial and the smoothing functions correspondingly.
The first one characterizes the dependence of the expected reward on the intended
saccade length. The motivation behind the introduction of the radial function
$R$ are both growing uncertainty of the fixation placement (\ref{eq:saccade-1})
and the duration of the step (\ref{eq:duration}) with the length of the saccade.
We assume that the radial function $R$ equals zero outside an interval $\left[a_{min},a_{max}\right]$,
where $a_{min}$ and $a_{max}$ are minimal and maximal saccade length
correspondingly. The minimal saccade length $a_{min}=1\deg$ \cite{otero2008saccades}
is chosen as a magnitude of the shortest possible voluntary movement.
The maximal saccade length $a_{max}=\sqrt{2}\cdot15\deg$ is equal
to the length of the diagonal of the stimulus image in our experiments. The
smoothing function $S$ describes the relative contribution of the surrounding
locations to the reward. The smoothing function has the same role as a term
$F$ (see eq. (\ref{eq:pi0})) in the definition of the information maximization
policy $\pi_{0}$, and it basically defines how meaningful the certain
location is without consideration of the time costs of a relocation. 

The form of policy (\ref{eq:generalform}) in the separable approximation
is:

\[\mu(D_{n},p_{n})\propto\]
\begin{equation}
\exp\left(R\left(\left\Vert D_{n}-A_{n}\right\Vert \right)\left(p_{n}*S\right)\left(D_{n}\right)\right)\label{eq:policyform}
\end{equation}
which is used in the simulation of the trajectories and the training phase. Two
heuristic policies presented in section \ref{sub:heuristics}
are both special cases of the general form of a policy in the separable approximation(\ref{eq:policyform}).

\subsection{Parametrization of policy} \label{sub:parametrixation}

The radial $R(x)$ and the smoothing $S(x)$ functions are represented
with Fourier-Bessel series:

\begin{equation}
R(x)=\begin{cases}
\begin{array}{c}
\sum_{\xi=1}^{\varXi}r_{\xi}J_{1}\left(\frac{u_{1:\xi}(x-a_{min})}{a_{max}-a_{min}}\right),\;a_{min}<x<a_{max}\\
0,\;\;\;\;\;\;\;\;\;else
\end{array}\end{cases}\label{eq:radial-1}
\end{equation}

\begin{equation}
S(x)=\begin{cases}
\begin{array}{c}
\sum_{\xi=1}^{\varXi}s_{\xi}J_{0}\left(\frac{u_{0:\xi}x}{b}\right),\;x<b\\
0,\;\;\;\;\;\;\;\;else
\end{array}\end{cases}\label{eq:smoothing-1}
\end{equation}
where $u_{i:\xi}$ are zeros of Bessel function of order i and $b$
is the radii of the visual field. This representation allows us to control
the dimensionality of the kernel and to effectively store the policy in
memory. The choice of orders ($i=0,1$) of Bessel functions in (\ref{eq:radial-1},\ref{eq:smoothing-1})
is caused by boundary conditions for the radial and the smoothing functions:
$R(a_{min})=R(a_{max})=0$; $S(b)=0$. The boundary conditions on the
radial function forbid the model observer to fixate the same location
again $R(0)=0$ and to make unlikely large saccades $R(a_{max})=0$.
The condition on the smoothing function $S(b)=0$ corresponds to the absence
of any information gain from remote locations, and, therefore, their
irrelevance to the process of fixation selection. So, the policy $\mu(\vartheta)$
is represented by set of parameters: $\vartheta\equiv\left(r_{0:\Xi},s_{0:\varXi}\right)$.

\subsection{REINFORCE}

We solve the optimization problem for the value function (\ref{eq:value_function})
with a policy gradient algorithm adopted from \cite{peters2006policy}.
This optimization procedure is represented as an iterative process of a
gradient estimation and an update of the policy parameters at the end of
each training epoch - the sequence of $M$ episodes. 
\begin{description}
\item [{Repeat}]~

\begin{enumerate}
\item Perform a training epoch with $M$ episodes and get the sequence of
observations, actions and costs for each time step $t$ and episode
$m$: $\left(p_{t,m},a_{t,m},V_{m}\right)$.
\item Estimate optimal baseline for each gradient element $\xi$: $b_{\xi}=\frac{\underset{m}{\sum}\left(\underset{t}{\sum}\nabla_{\xi}\log\mu_{\vartheta}\left(a_{t,m},p_{t,m}\right)\right)^{2}V_{m}}{\underset{m}{\sum}\left(\underset{t}{\sum}\nabla_{\xi}\log\mu_{\vartheta}\left(a_{t,m},p_{t,m}\right)\right)^{2}}$
\item Estimate the gradient for each element: $\eta_{\xi}=\underset{m}{\sum}\left(\underset{t}{\sum}\nabla_{\xi}\log\mu_{\vartheta}\left(a_{t,m},p_{t,m}\right)\right)^{2}\left(b_{\xi}-V_{m}\right)$
\item Update policy parameters: $\vartheta\leftarrow\vartheta+\alpha\eta$
\end{enumerate}
\item [{until}] gradient $\eta$ converges. 
\end{description}

\subsection{PGPE}

The second approach to the optimization problem (\ref{eq:value_function})
is a parameter exploring policy gradient presented in \cite{sehnke2010parameter}.
As well as in the previous section, we estimate the gradient and update
the policy parameter at the end of each training epoch. We use a symmetric
sampling of the policy parameters for gradient estimation. At the beginning
of each step we generate the perturbation $\epsilon$ from normal
distribution $N(\mathbf{0},\boldsymbol{I}\mathbf{\boldsymbol{\sigma}}^{2})$
and create the symmetric parameter samples $\vartheta^{+}=\mu+\epsilon$
and $\vartheta^{+}=\mu-\epsilon$, where $\mu$ is the current values of the
policy parameters for the training epoch. Then we simulate one episode
for each parameter sample and denote the cost $V^{+}$ for the episode
generated with $\vartheta^{+}$ , and $V^{-}$ for $\vartheta^{-}$
correspondingly. At the end of each training epoch the policy parameters
and the standard deviation of the distribution of perturbation are updated according
to the equations:

\begin{equation}
\mu_{i}=\mu_{i}+\alpha\sum_{j=1}^{M}\epsilon_{j}^{i}\left(V_{j}^{-}-V_{j}^{+}\right)
\end{equation}

\begin{equation}
\sigma_{i}=\sigma_{i}+\alpha\sum_{j=1}^{M}\left(\frac{\left(\epsilon_{j}^{i}\right)^{2}-\sigma_{i}^{2}}{\sigma_{i}}\right)\left(\left\langle V\right\rangle -\frac{V_{j}^{+}+V_{j}^{-}}{2}\right)
\end{equation}
where for $j$th episode $\epsilon_{j}^{i}$ is the perturbation for the parameter
$i$ and $V_{j}^{\pm}$ are sampled costs. The cost baseline is
chosen as a mean cost for the training epoch.

\subsubsection{Convergence of policy gradient\label{sub:Convergence-of-policy-1}}

The Markov decision process defined by set of dynamic equations (\ref{eq:observation},\ref{eq:saccade-1},\ref{eq:duration},\ref{eq:policyform})
was simulated on $N\times N$ grid, which comprises the $N^{2}$ possible
target locations, where $N$=128. At the beginning of the optimization
procedure we pick the policy parameters $\vartheta$ randomly from the
uniform distribution $U(-0.5,0.5)$ and fix parameter $\lambda=0.001$.
For both algorithms we use the same parametrization of policy. The
training epoch for both PGPE and REINFORCE consists of 400 episodes.
Learning rate $\alpha=0.2$ was the same for both algorithms. 

\begin{figure}
\includegraphics[scale=0.72]{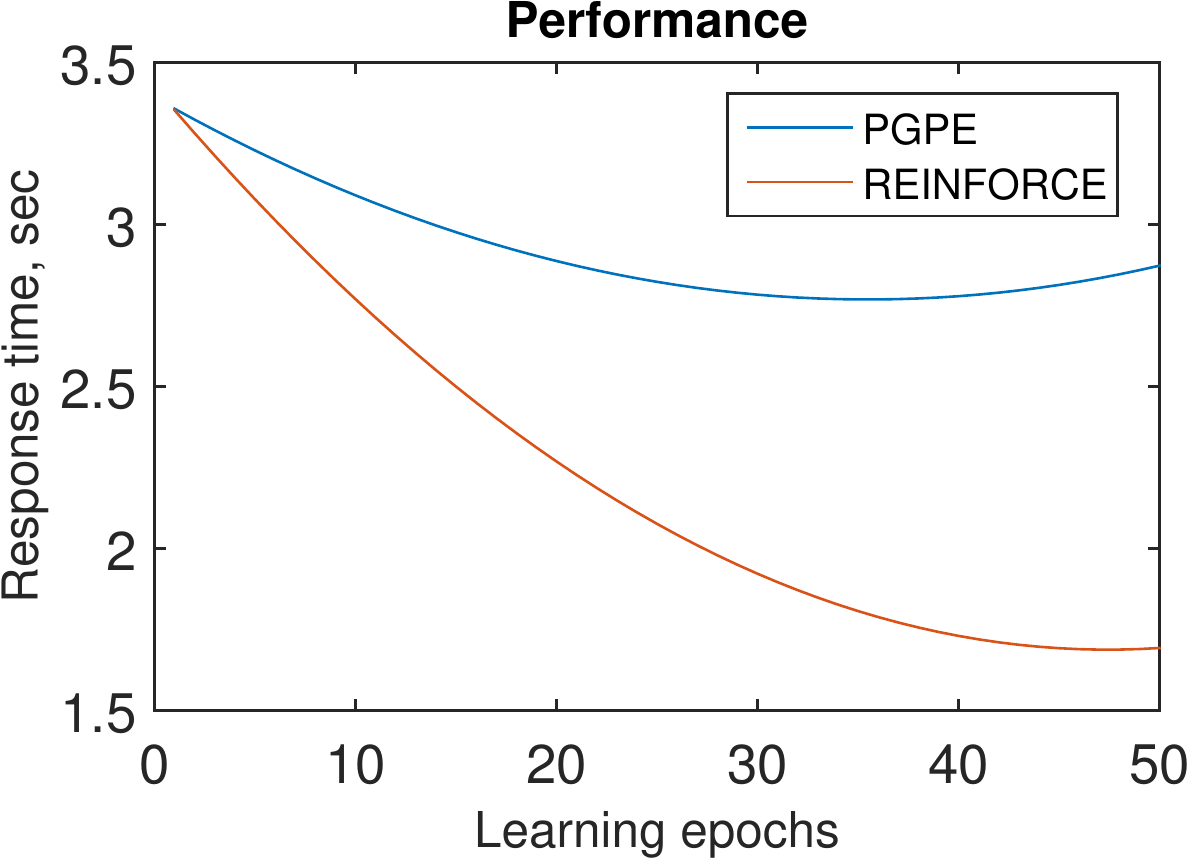}

\caption{\label{fig:Performance-of-parameter-1}The performance of parameter exploration
policy gradient (PGPE) and episodic REINFORCE with an optimal baseline.}
\end{figure}

Figure \ref{fig:Performance-of-parameter-1} illustrates the performance
of two policy gradient methods we used for search of the optimal policy
for the case of FPOC corresponding to $e_{n}=0.25$ and $e_{t}=0.2$.
Both algorithms used Fourier-Bessel parametrization of policy with a
dimensionality $\varXi=45$ for the radial and the smoothing functions. REINFORCE
performed better for all parameter settings. On average, it takes
around 50 and 40 learning epochs to converge for REINFORCE and PGPE
correspondingly. The choice of the dimensionality higher than 45 doesn't
improve the performance of both algorithms. 

\begin{figure*}
\includegraphics[scale=0.58]{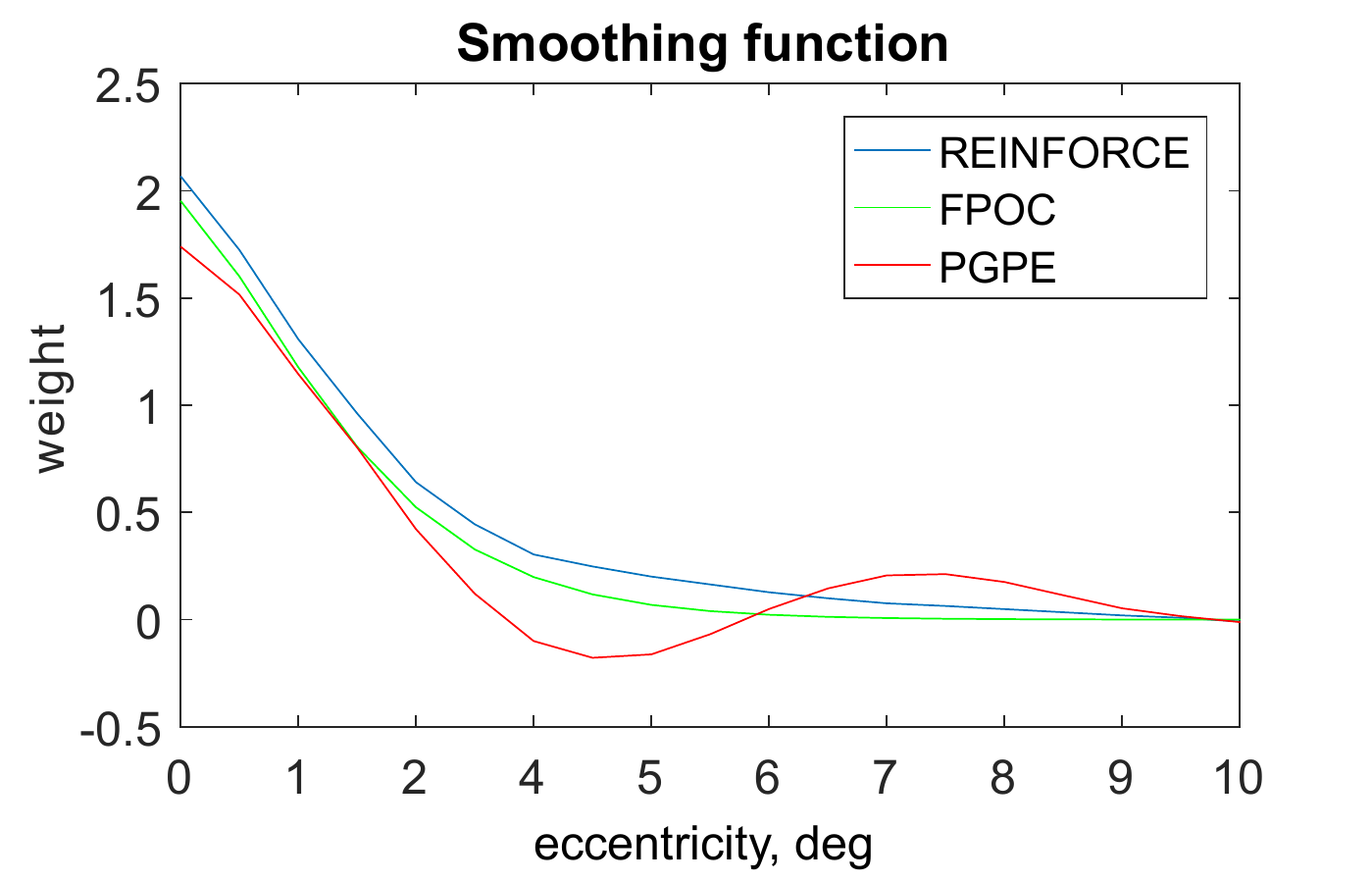}\includegraphics[scale=0.58]{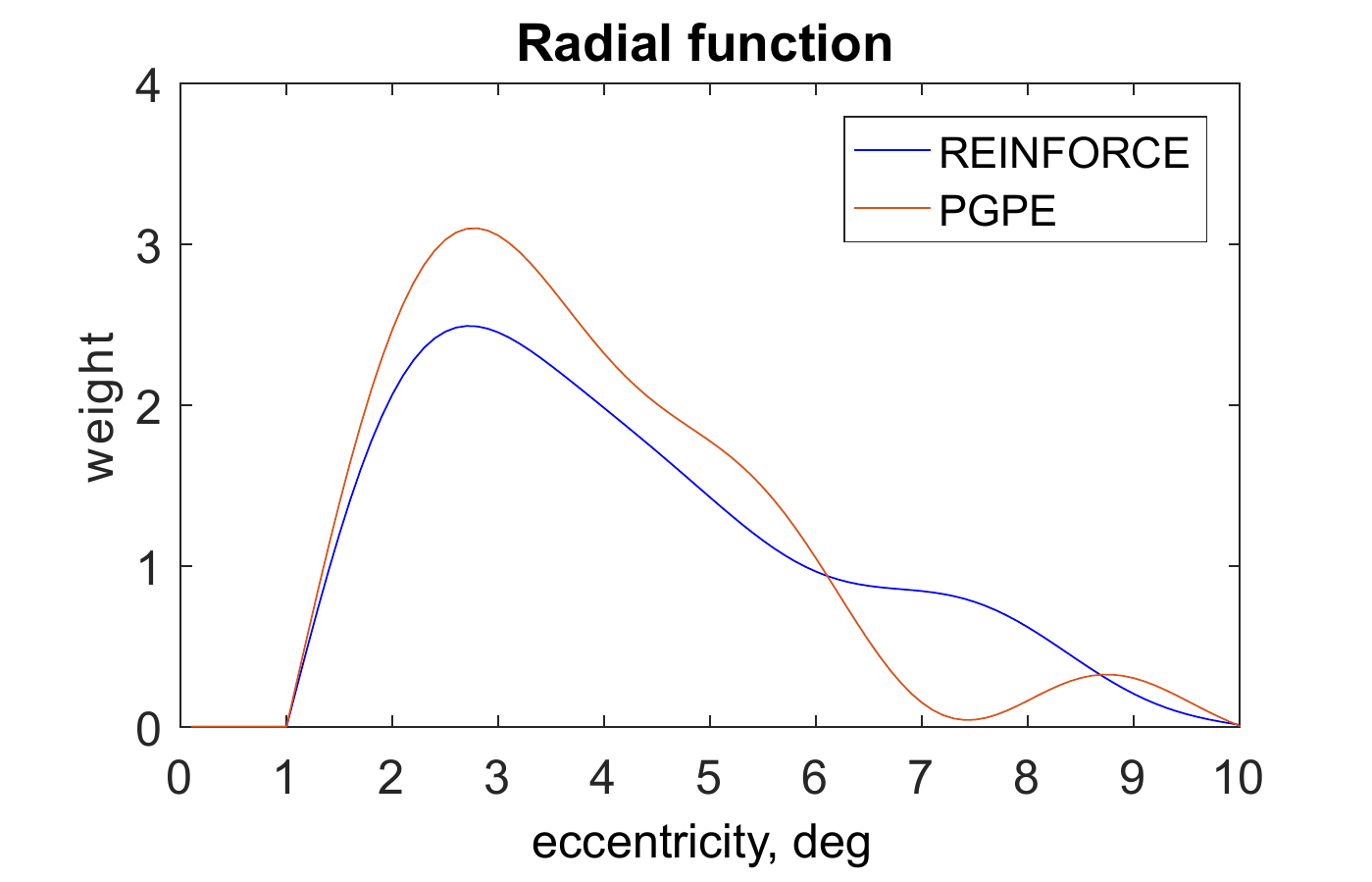}\caption{\label{fig:Results-of-optimization:-1}The results of the optimization: the smoothing S(x) funtion (left) and the radial
R(x) function (right).}
\end{figure*}

Figure \ref{fig:Results-of-optimization:-1} shows the results of
optimization: the radial $R(x)$ and the smoothing $S(x)$ functions. Both
REINFORCE and PGPE provide close estimates of the smoothing and the radial
functions for eccentricity smaller than $\epsilon<3\textdegree$.
In order to compare the solution with the heuristic policies (\ref{eq:pi0},\ref{eq:pi_0-1}),
we presented FPOC on the same plot with the smoothing function. The smoothing
function provided by REINFORCE is monotonously decreasing as well
as FPOC, whereas for PGPE we have a fluctuating solution with a decreasing
amplitude of oscillations. The behavior of the radial function is similar
for both solutions, with higher amplitude of oscillations for PGPE
solution. 

\section{Implementation of MF-DFA\label{sec:Multifractal-analysis}}

\subsection{\label{sub:Multifractal-analysis-intro}Multifractal analysis}

In this chapter we present the 
details of MF-DFA algorithm used here for calculation of the generalized
Hurst exponent. All of this section is based on Kantelhardt et al.\cite{kantelhardt2002multifractal}.

The procedure of MF-DFA starts with definition of a profile for time
series $X=\left\{ x_{1},...,x_{N}\right\} $ with a compact support:

\begin{equation}
Y(i)=\sum_{k=1}^{i}\left(x_{k}-\left\langle x\right\rangle \right)
\end{equation}

The profile $Y(i)$ is divided on $N_{s}\equiv\mathrm{int}(N/s)$
segments, where $s$ is chosen among some linear space $s\in S=\left\{ s_{min,}s_{min}+\triangle s,...,s_{max}\right\} $.
The segmentation starts from the beginning of time series, therefore, there are residual $N\div s$ number of the elements at the end of time-series.
In order to process the residual elements, the segmentation is also
performed from the end of time series. So, at the end of segmentation
procedure we have $2N_{s}$ segments for each value of $s$.

The calculation of the variance is based on an approximation of local trend
for each segment $\nu=1,...,N_{s}$ with a polynomial function $y_{v}$.
Then, the variance on each segment is calculated as:

\begin{equation}
F^{2}(\nu,s)=\frac{1}{s}\sum_{i=1}^{s}\left\{ Y\left[\left(\nu-1\right)s+i\right]-y_{v}(i)\right\} 
\end{equation}

for each segment $\nu=1,...,N_{s}$ and 

\begin{equation}
F^{2}(\nu,s)=\frac{1}{s}\sum_{i=1}^{s}\left\{ Y\left[N-\left(\nu-N_{s}\right)s+i\right]-y_{v}(i)\right\} ^{2}
\end{equation}

for $\nu=N_{s}+1,...,2N_{s}$. The order m of polynomial function
must satisfy the condition $m\leq s-2$. The variance over all segments
are averaged to obtain the $q$th order fluctuation function:

\begin{equation}
F_{q}(s)=\left\{ \frac{1}{2N_{s}}\sum_{\nu=1}^{2N_{s}}\left[F^{2}(\nu,s)\right]^{q/2}\right\} ^{1/q}\label{eq:fluctuation_function}
\end{equation}

According to the properties of $q$th order fluctuation function \cite{peitgen2006chaos},
the scaling behavior of $F_{q}(s)$ is governed by the generalized Hurst
exponent:

\begin{align}
F_{q}(s)\sim s^{H(q)}\label{eq:scaling}
\end{align}

The value of $H(q)$ is usually obtained through a linear regression
of $\log_{2}\left(F_{q}(s)\right)$.

\subsection{Interpolation of simulated trajectories\label{sub:Interpolation-of-sequence}}

We perform MF-DFA analysis on the magnitude of the saccadic events simulated
by MDP defined above. Each episode of MDP provides the sequence of
vectors of gaze positions: $\mathbf{A}_{1},...,\mathbf{A}_{N}$. In
order to get the time series of the gaze allocation in real time - $\bar{A}$,
we follow the simple procedure of an interpolation:
\begin{itemize}
\item The calculation of the duration of each time step $n$ with \ref{eq:duration},
the total time of episode $T=\sum_{i=1}^{N}\Theta_{i}$ and start
time of each discrete step:
\begin{equation}
T_{n}=\left\{ \begin{array}{c}
0,\,if\,n=1\\
\sum_{i=1}^{n-1}\Theta_{i},\,if\,n>1
\end{array}\right.
\end{equation}
 
\item The choice of the length of the real-time sequence $M=25*T$, which corresponds
to 40 millisecond resolution.\label{Choice-of-length-1}
\item For each element $t$ of $\mathbf{\bar{A}}$ we define, which discrete
time step it belongs: $T_{n}\leq\frac{t}{25}<T_{n+1}$. 
\item If time step $t$ of $\mathbf{\bar{A}}_{t}$ corresponds to the fixation
during the discrete time step $n:$ $\frac{t}{25}-T_{n}<\varTheta_{fix}(n)$,
than $\mathbf{\bar{A}}_{t}=\mathbf{A}_{n}$. In the other case, if
time $t$ corresponds to the saccadic movement within the discrete time interval
n, we have: $\mathbf{\bar{A}}_{t}=\mathbf{A}_{n}+\tau_{sac}\frac{\mathbf{A}_{n+1}-\mathbf{A}_{n}}{\left|\mathbf{\mathbf{A}_{n+1}-\mathbf{A}_{n}}\right|}\left(\frac{t}{25}-T_{n}-\varTheta_{fix}(n)\right)$.
Therefore, we have defined the function that maps the discrete sequence
$A$ to real-time sequence $\bar{A}$.
\end{itemize}
The real-time sequences $\bar{A}$ from 1000 episode corresponding
to each policy are merged, and the resulting sequences $\bar{A}_{\mu},\bar{A}_{\pi1},\bar{A}_{\pi0}$
are analyzed with MF-DFA. 
\subsection{Multifractality of simulated trajectories\label{sub:mf_simulated}}

We perform MF-DFA over the differentiated trajectories generated with
PO-MDP under the heuristic policies $\pi_{0},\pi_{1}$ and the learned policy
$\mu$. Before differentiation trajectories were represented as real
time sequences with the procedure of the interpolation \ref{sub:Interpolation-of-sequence}.

The model presented here is not devoted to FEM and can't describe
the combined movement of both FEM and the saccades. The results of our
analysis should be compared with the scaling behavior of $F_{q}(s)$
for human eye-movements on the scales $s\geq256$ $ms$, which corresponds
to the upper regime. Therefore, we set the minimal time scale $s_{min}=256 ms$. The choice of $s_{max}=2*10^{3}ms$ corresponds
to the average length of episode. We assume that there is no correlation
between the episodes due to a random location of the first fixation and the location
of the target. 

\begin{figure*}

\includegraphics[scale=0.70]{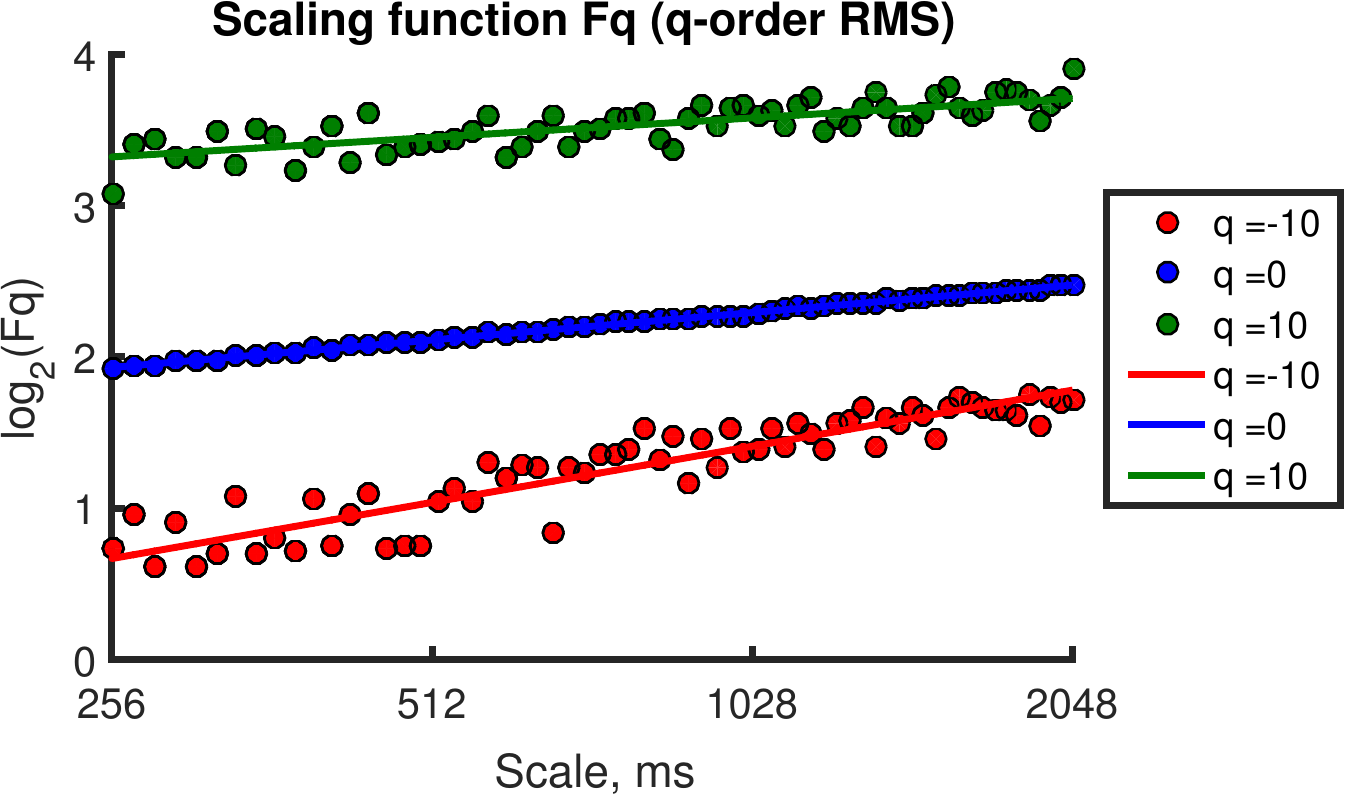}\includegraphics[scale=0.75]{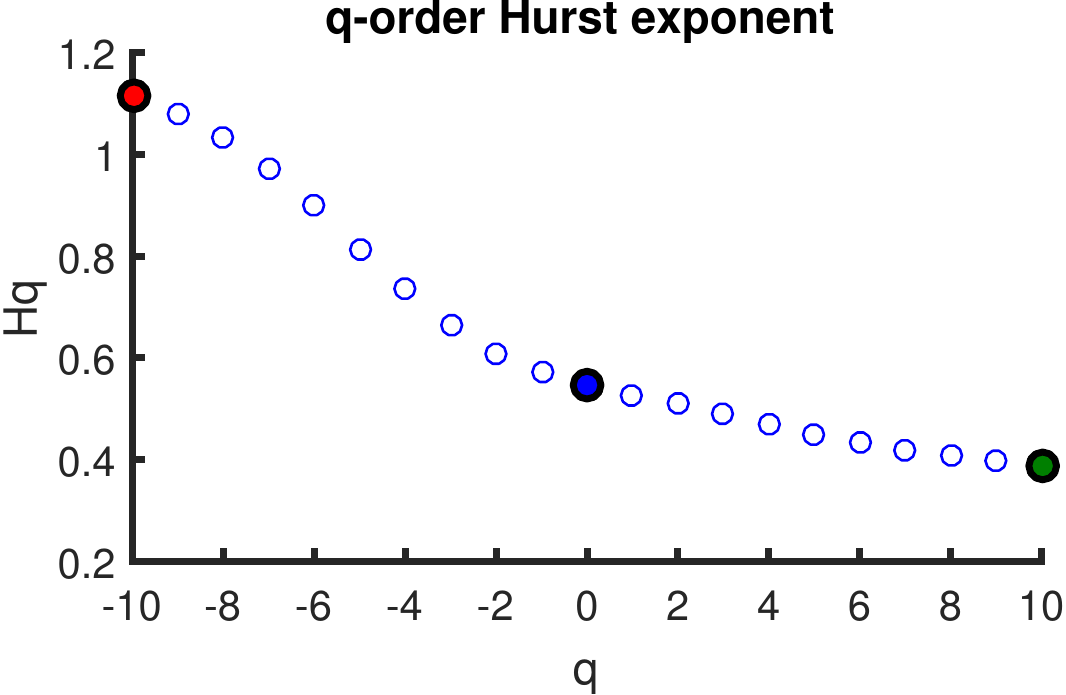}\caption{\label{fig:Fluctuation function simulated}The scaling of the q-order
fluctuation function $F_{q}(s)$ (left) estimated for the simulated trajectories
under the infomax rate policy $\pi_{1}$, and the generalized Hurst exponent
$H(q)$ (right) computed through a linear regression of $\log_{2}\left(F_{q}(s)\right)$.
The red, blue and green lines correspond to the linear approximation of
function $\log_{2}\left(F_{q}(s)\right)$ for orders $q=\left\{ -10;0;10\right\} $.
The scaling of $F_{q}(s)$ doesn't exhibit the crossover for the positive
q-orders on an interval of scales $\left[s_{min,}s_{max}\right]$.}

\end{figure*}

The figure \ref{fig:Fluctuation function simulated} demonstrates the
scaling of the q-order fluctuation function $F_{q}(s)$ (\ref{eq:fluctuation_function})
for simulated trajectory under the infomax greedy policy $\pi_{0}$ for
the conditions: $e_{t}=0.2$, $e_{n}=0.15$. The red, blue and green lines
correspond to the linear approximation of the function $\log_{2}\left(F_{q}(s)\right)$
for the orders $q=\left\{ -10;0;10\right\} $. The scaling of $F_{q}(s)$
doesn't exhibit the crossover for positive q-orders on an interval of
scales $\left[s_{min,}s_{max}\right]$, however the behavior of $\log_{2}\left(F_{q}(s)\right)$
deviates from linear at the large scales $s\sim s_{max}$. The simulations
on different grid sizes, which correspond to different average time
of task execution, have shown that the interval of linear behavior
of $\log_{2}\left(F_{q}(s)\right)$ always coincides with $\left[s_{min,}s_{max}\right]$. The scaling of $F_{q}(s)$
on $\left[s_{min,}s_{max}\right]$ is different for different orders
$q$ and, therefore, the trajectories $\bar{A}_{\pi_{1}}$ are multifractal
time series.

\begin{figure*}

\includegraphics[scale=0.64]{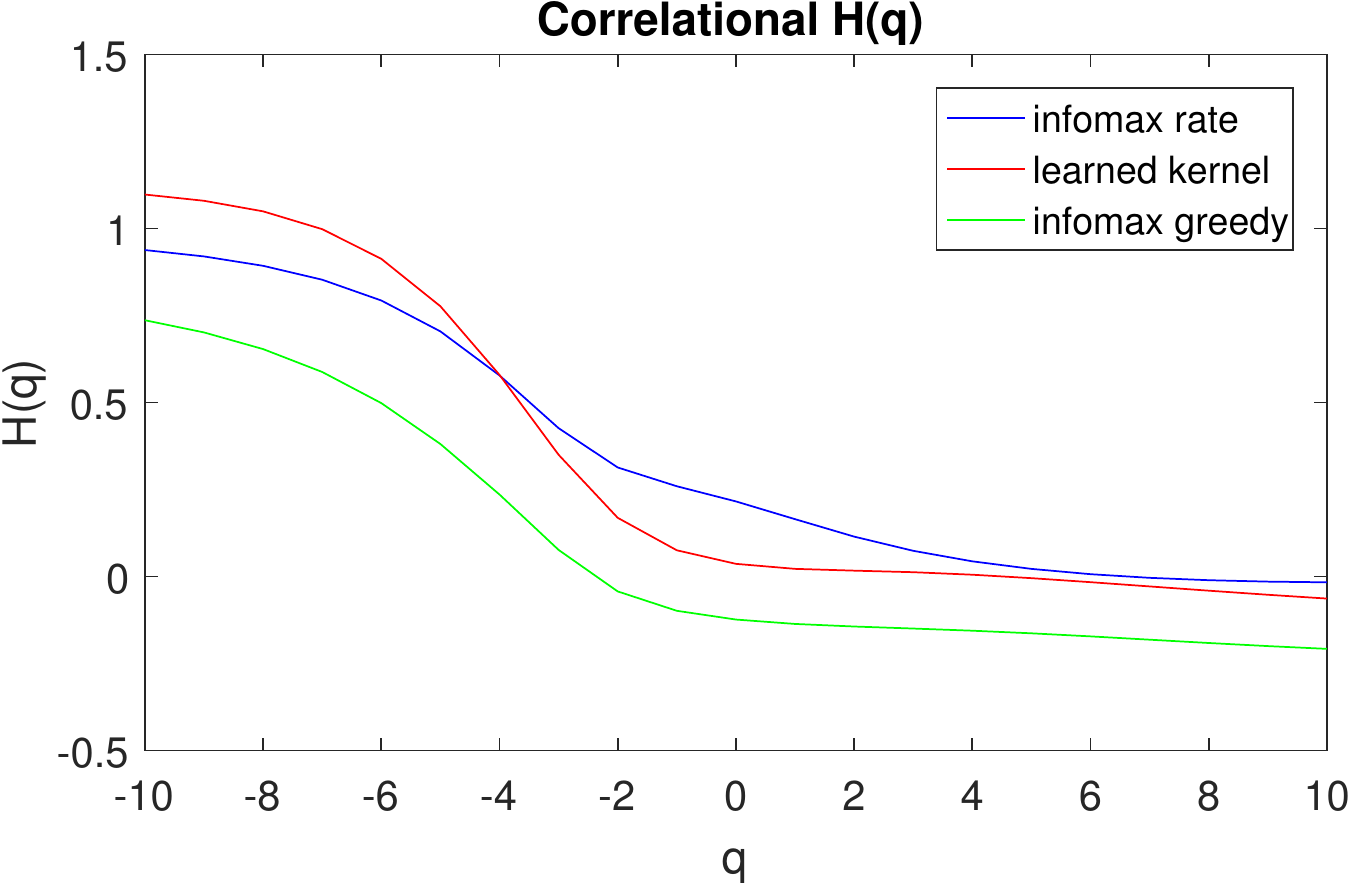}\includegraphics[scale=0.64]{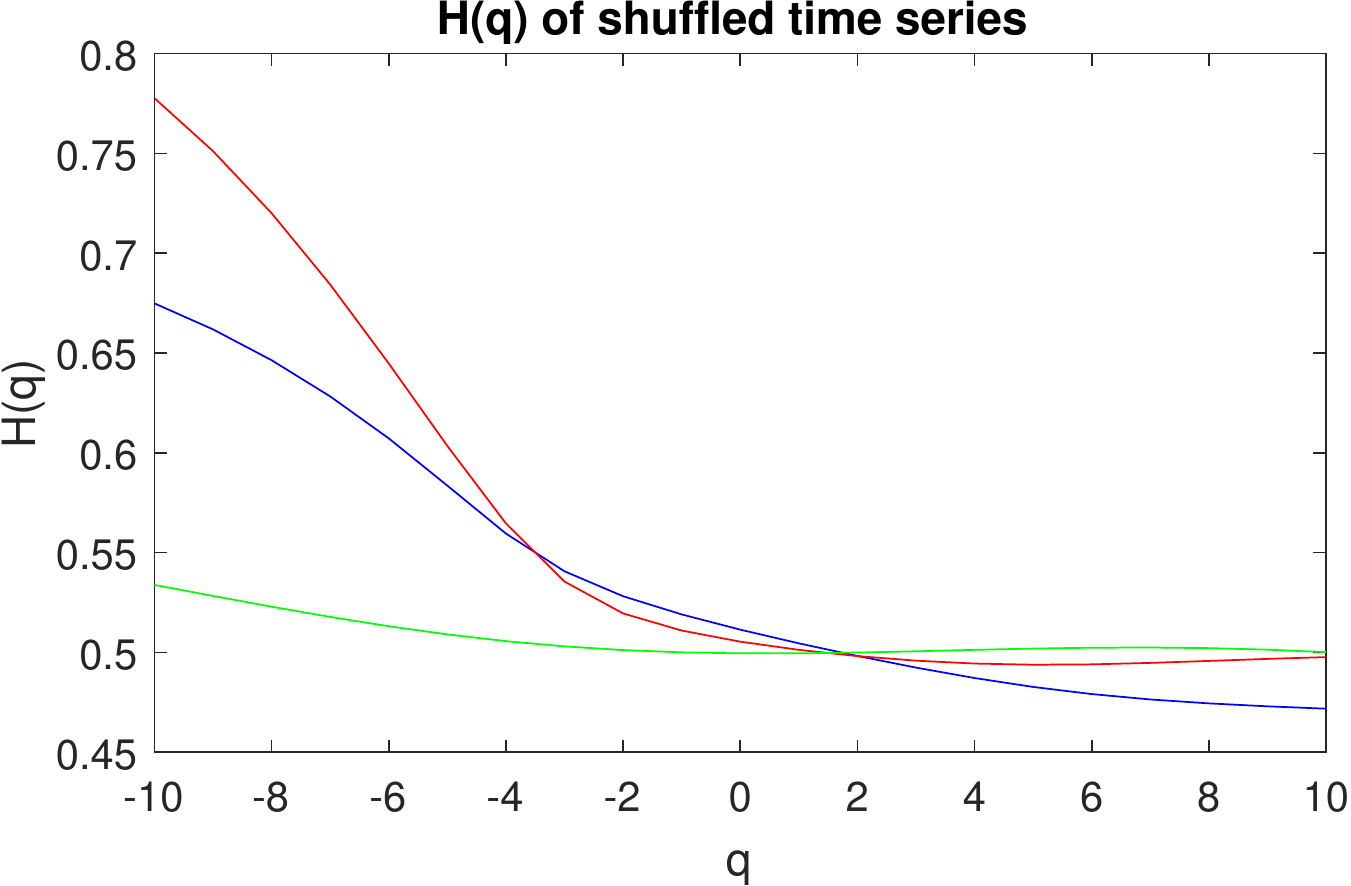}\caption{\label{fig:Hurst-exponent-of-sim}The Hurst exponent of the shuffled time
series $H_{shuf}(q)$ (right) and the correlational Hurst exponent $H_{corr}(q)$
(left) of the trajectories simulated under the different policies. As well
as in the case of human eye-movements, two types of multifractality
present in the simulated time series. The
behavior of $H_{shuf}(q)$ resembles the power-law distribution scenario
\ref{eq:Levy} for all policies, except the infomax greedy $\pi_{0}$.  The distribution of saccade length doesn't correspond to the power-law for   $\pi_{0}$ , which was demonstrated on the figure \ref{fig:Length-distribution-of} . On the contrary, the infomax rate $\pi_{1}$ and the learned policy $\mu$  generate the movement with the distributional multifractality that presents in human eye-movements as well. }

\end{figure*}

The figure \ref{fig:Hurst-exponent-of-sim} demonstrates our estimates
of the correlational Hurst exponent $H_{corr}(q)$ (left) and the Hurst exponent
of shuffled time series $H_{shuf}(q)$ of time series simulated under
the different policies. As well as in the case of the human eye-movements,
two types of multifractality present in the simulated time series. The
behavior of $H_{shuf}(q)$ resembles the power-law distribution scenario
\ref{eq:Levy} for all policies, except the infomax greedy $\pi_{0}$.  The distribution of the saccade length doesn't correspond to the power-law for   $\pi_{0}$ , which was demonstrated on figure \ref{fig:Length-distribution-of} . On the contrary, the infomax rate $\pi_{1}$ and the learned policy $\mu$  generate the movement with the distributional multifractality that presents in human eye-movements as well. 

For all policies the correlational
Hurst exponent is positive for the negative $q$-orders. This indicates
the presence of long-range correlations for small fluctuations.  The large fluctuations are anticorrelated
for $\pi_{1}$ and exhibit weak anti-correlation for $\mu$. We observe
the last scenario for upper regime of human eye-movements \ref{fig:Correlational-generalized-Hurst},
where the large fluctuations demonstrate a weak anticorrelation in a contrary
to positively correlated small fluctuations.

\section{Implementation of psychophysical experiments \label{sec:Qualitative-analysis}}

We set a goal to reproduce the eye tracking experiment described in
\cite{najemnik2005optimal}. In this section we provide the description
of the psychophysical experiments.

\subsection{Participants}

The group of nine patients with normal to corrected-to-normal vision
participated in the experiment. The group included four postgraduate
students (age $23\pm7$, 4 males) from Queen Mary University of London. This group was aware of the experimental settings and passed 10 minutes
of training sessions with four different experimental conditions, which correspond to the certain value of the RMS contrast of background noise. The experiments were approved
by the ethics committee of Queen Mary University of London and informed
consent was obtained.
\subsection{Equipment}

We used DELL P2210 22'' LCD monitor (resolution $1680\times1050$,
refresh rate 60 Hz) driven by a Dell Precision laptop for all experiments.
The eye movements of the right eye were registered using Eye Tracker device
SMI-500 with a sampling frequency of $120$ Hz. The Eye tracker device
was mounted on the monitor. Matlab Psychtoolbox was used to run the experiments
and generate the stimulus images.

\subsection{Stimulus and procedure}

The participants set in front of the monitor with their heads fixed with a
chin rest at a distance of 110 $cm$ from the monitor. The monitor subtended
a visual angle of $21\times15\deg$. Each participant was shown
the examples of the stimulus image before the experiments and was instructed
to fixate the target object as fast as possible and to press the certain
button on a keyboard to indicate that they found the target. All four participants completed
one practice session with 40 trials before the experiment. 

The stimuli were the static images generated before each session according
to the description from the original experiment \cite{najemnik2005optimal}.  The 1/f noise was generated on a square
region on the screen, which spans the visual angle of
$15\times15 \deg$. The target was 
sine grating $6\deg^{-1}$
framed by symmetric raised cosine. The target appeared randomly at
any possible location on the stimuli image within the square region. The
experiments were provided for one level of RMS contrast of target
$e_{t}=0.2$ and several levels of 1/f noise RMS contrast $e_{n}\in(0.1,0.15,0.2,0.25)$.

The participants completed four experimental sessions with 120 trials.
The experimental session started after inbuilt nine-point grid calibration
of the eye-tracking device. The participants were given 3 minutes of rest
between sessions. One of 120 stimuli images were shown at the beginning
of each trial. The participants are assumed to perform the visual search
task, which is finished by pressing the "END" button. In our experimental
settings, the signal from the participants was blocked for $300$ $ms$
from the start of each trial. If the gaze position measured by the eye
tracking device is in the vicinity of $2$ $\deg$ around the location of the target
at the moment participant presses the "END" button, the task is
considered successful. Due to the presence of a temporal delay between the moments of localization of target and pushing of "END" button we block the signal from END button for $400$ $ms$. After a completion of each trial the central
fixation cross was shown for $500$ $ms$, then the next trial started
and new stimulus image was shown to participants. 
\begin{figure*}
\includegraphics[scale=0.35]{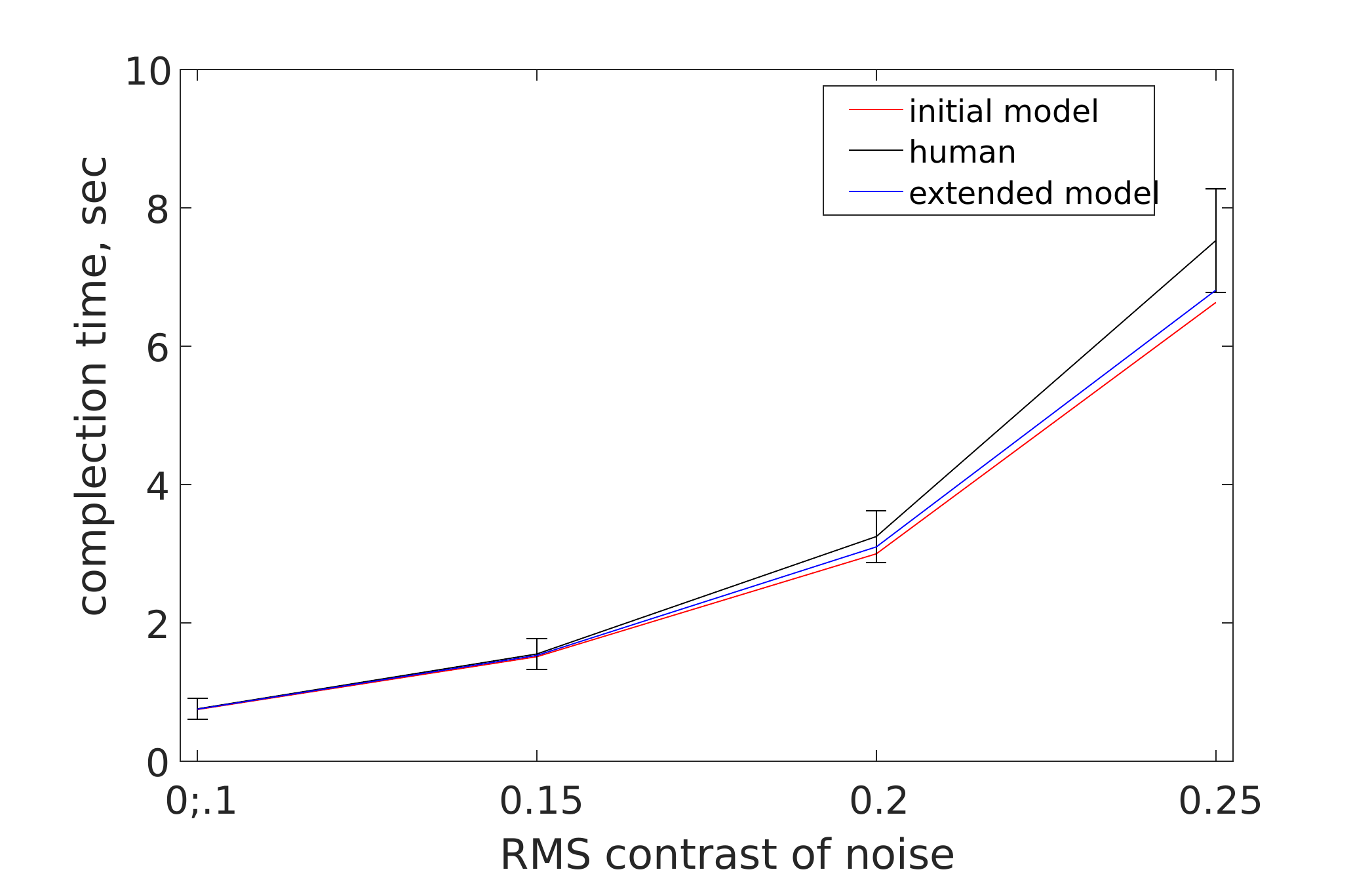}
\includegraphics[scale=0.35]{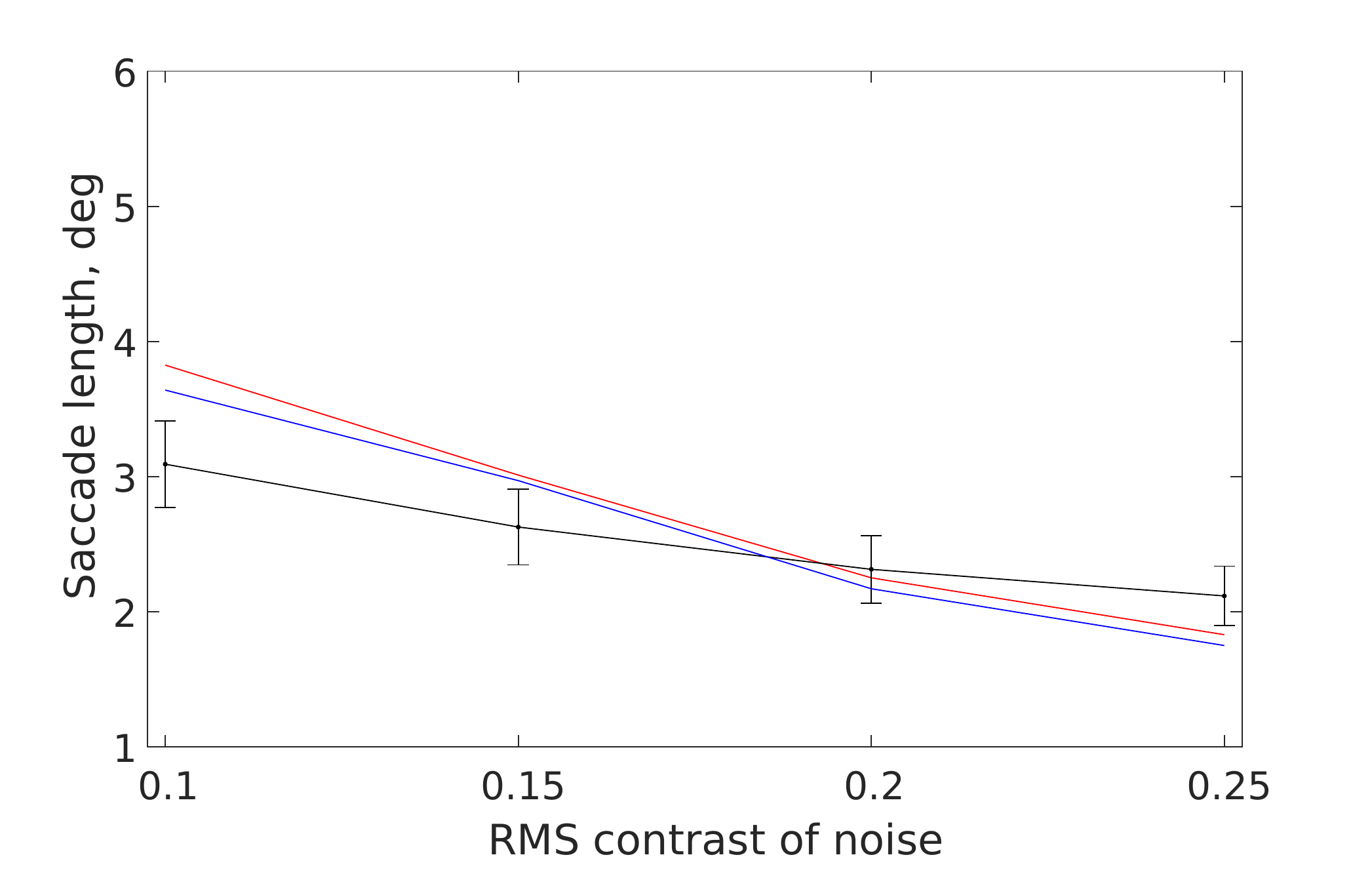}
\caption{The completion time (left) and the saccade length (right) for the human observers, the simulated agents with the initial observation  model and the extended one. }
\label{fig:comparison}
\end{figure*}

\section{Influence of saccade latency} \label{sec:latency}
According to the literature, the saccade programming is assumed to be the two-stage process that consists of labile and non-labile stages \cite{engbert2005swift}. The labile stage is the first stage of the saccade programming, during which the initial saccade command can be cancelled in a favour of saccade to another location. The saccade to the next location is executed after the non-labile stage. The visual input is active during both labile and non-labile stages and suppressed during the execution of a saccade. Therefore, the decision $D_{n}$ is made at the end of the labile stage, and the visual input received at the location $A_{n}$ during non-labile stage of saccade programming can be used for decision-making only at the next step $n+1$.  As a result, the observer receives two separate observation vectors $\mathbf{W^{n+1}_{lb}}$ and   $\mathbf{W^{n+1}_{nlb}}$  from the previous and the current fixation locations. The observation model \ref{eq:observation} doesn't take into account the duration of the observation. We assume that observation vector is integrated continuous-time Gaussian white noise: $W(l,A)=\int_{t}^{t+\varTheta} w(t,l,A) dt $, which satisfies following: 

\begin{enumerate}
\item $E[w(t,l,A) ]=\delta_{l,m}/\theta_{0}$ \label{eq:cn1}
\item $E[w(t_{1},l_{1},A) ,w(t_{2},l_{2},A) ]=\frac{\delta_{t_{1},t_{2}}\delta_{l_{1},l_{2}}  }{\theta_{0}   F^{2}\left(\left\Vert l-A\right\Vert \right)}$ \label{eq:cn2}
\end{enumerate}
where $\theta_{0}=250$ $ms$ is a time interval is the detection experiment \cite{najemnik2005optimal}, for which the visibility maps were measured. This model of noise generalizes the "noisy observation" paradigm for variable fixation duration. The result of integration of continuous time noise $w$ is Gaussian white noise $W$ with mean  $\varTheta \cdot  E[w]$ and variance $\varTheta \cdot \sigma^{2}[w]$. Next, we assume that the duration of  the non-labile stage is  $\varTheta_{nlb}= 41,6$ $ms$ \cite{engbert2002dynamical} and the rest of fixation duration is allocated for the labile stage $\varTheta_{lb} \approx 200$ $ms$ (the average fixation duration according to our data: $\varTheta_{fix}=240$ ms).  Using \ref{eq:cn1}  and \ref{eq:cn2} we compute mean and variance of observation inputs   $\mathbf{W^{n+1}_{lb}}$ and   $\mathbf{W^{n+1}_{nlb}}$ , and successively apply the equation \ref{eq:inference} to evaluate the belief state $p_{n+1}$. 
\par We learned the policy of gaze allocation for the extended observation model using REINFORCE.   We compared the basic characteristic of trajectories simulated under this policy with the simulations for the initial model and data from the human observers (Look at \ref{fig:comparison}.  The initial model outperformed the extended one, but no significant difference was found. We didn't expect any significant difference in performance, because the observer receives the same amount of information on average  in both models.



%

\section*{References}

\bibliographystyle{elsarticle-num-names}
\addcontentsline{toc}{section}{\refname}\bibliography{proplib}

%




\end{document}